# Decoding Composition–Structure–Performance Relations in Sodium Layered Oxide Positive Electrode Materials

Yunyeong Choi[1,2], Hasindhi H. Lokuge[1,3], Joshua D. Bocarsly[1,3], and Pieremanuele Canepa[1,2,4*]

[1]Texas Center for Superconductivity at the University of Houston, Houston, TX 77204, USA
[2]Department of Electrical and Computer Engineering, University of Houston, Houston, TX 77204, USA
[3]Department of Chemistry, University of Houston, Houston, TX 77204, USA
[4]Department of Mechanical Engineering, Purdue University, West Lafayette, IN 47907, USA

*e-mail: canepa@purdue.edu

## Abstract

Layered sodium transition-metal (TM) oxides are leading sodium-ion cathode materials, yet their Na/vacancy orderings, stacking sequences, and cooperative Jahn–Teller (JT) distortions are known for only a handful of compositions. We compile 41 unique Na/vacancy orderings reported previously into a single library and apply it uniformly to five single-TM and all ten $Na_xTM_{0.5}TM'_{0.5}O_2$ oxides (TM = Ti, Mn, Fe, Co, Ni), screening over 120,000 structures with machine-learning interatomic potentials before 1,025 first-principles calculations. Phase diagrams explain the observed O3/P3/O1 sequences and voltages; the O3-to-P3 competition hinges on ordering energies of a few meV/atom that scalar descriptors, such as the cationic potential and the interlayer-slab distance, miss. The redox couples form a nearly partner-independent ladder, $Ti^{3+/4+}$ < $Mn^{3+/4+}$ < $Co^{3+/4+}$ < $Ni^{2+/3+}$ < $Fe^{3+/4+}$, which explains pristine charge transfer and the oxidation sequence. Van Vleck analysis shows that TM mixing suppresses the cooperative JT distortions of $Mn^{3+}$ and $Ni^{3+}$ through charge transfer to JT-inactive states or dilution of JT-active centers. An energy-density-versus-strain map yields a design rule: a high-voltage couple without JT strain, realized by $Mn^{4+}/Ni^{2+}$charge transfer in $Na_xMn_{0.5}Ni_{0.5}O_2$, the most practical composition. We release the ordering-resolved dataset to train accurate machine-learning interatomic potentials for exploring higher compositional complexities in these oxides.

## Introduction

Sodium (Na)-ion batteries (NIBs) are a leading low-cost complement to lithium (Li)-ion batteries for stationary energy storage and entry-level mobility. Among NIB positive-electrode materials (hereafter referred to as cathode materials), polyanionic compounds—including phosphates, such as $Na_3V_2(PO_4)_3$ [1–5], and $Na_3V_2(PO_4)_2F_3$ [6–8]—as well as Prussian blue analogs[9,10] and layered transition-metal (TM) oxides, $Na_xTMO_2$, are prominent candidates. $Na_xTMO_2$ stand out for their high gravimetric and volumetric capacities, high operating voltages, good processability, and compatibility with established lithium-ion-battery electrode manufacturing platforms[11–16].

Selecting transition-metal compositions that deliver both high energy density and robust capacity retention still relies heavily on chemical intuition, and discovering improved cathode materials remains an active area of research[11–13,17–21]. In particular, the P2-, O3-, and P2/O3 hybrid-type stackings (or P2-, O3-phases) in Delmas notation[22] are among the most promising material families[12,17]. As illustrated in **Figure 1**, P and O denote prismatic or octahedral Na coordination between the $TMO_2$ slabs, and the digits 2 and 3 indicate the number of $TMO_2$ slabs in the repeat unit[22]. In general, P2-phases provide higher sodium-ion mobility, whereas O3-phases offer higher sodium density, reflected in higher gravimetric capacities due to their complete sodiation[12,17].

Unlike their lithium counterparts, layered $Na_xTMO_2$ often undergo reversible stacking transformations during sodium cycling, such as O3–P3 or P2–O2 transitions[17]. These transformations arise because the larger Na ions (compared to Li) can arrange into both octahedral and prismatic coordination sites, while gliding of the $TMO_2$ slabs interconverts these Na environments. In $TMO_2$, transition-metal-layer gliding is primarily governed by the degree of sodiation and is coupled to Na/vacancy (Na/Vac) ordering, transition-metal charge ordering within the TM layers, and lattice distortions, such as Jahn–Teller (JT) distortions of the $TMO_6$ octahedra[23–29]. The resulting highly "stepped" voltage profiles and anisotropic strain during desodiation drive capacity fade. Thus, these phase transitions or anisotropic strain need to be mitigated for commercial use[22,30–33]. Because TM composition governs both the stacking sequence and Na/Vac ordering, rational TM selection requires identifying the stable stacking arrangements and ordering patterns at each Na composition.

Previous experimental and theoretical studies of O3 phases have resolved Na/Vac orderings of a handful of single-transition-metal layered materials, including $Na_xVO_2$, $Na_xCrO_2$, $Na_xMnO_2$, $Na_xCoO_2$, and $Na_xNiO_2$ [23,26,27,29,34–37]. In-depth experiments and simulations often report Na/Vac orderings for specific compositions and conditions, making results hard to compare or transfer and complicating hypothesis development for improving these materials[23,35–40]. While detailed structural studies of Na/Vac ordering remain essential for designing improved layered-oxide cathode materials, the available data remain highly fragmented across the literature, underscoring the need for a comprehensive and systematic assessment. Furthermore, mixed-TM O3 compounds, which make up most practical NIB cathode materials, have been characterized electrochemically, but to our knowledge their Na/Vac orderings have never been computationally enumerated within a single consistent framework[41,42]. This study addresses these limitations through a systematic, comparative assessment of Na/vacancy ordering across layered sodium transition-metal oxide cathodes and its influence on the materials' electrochemical behavior.

To address this gap, we consider the three stackings defined in **Figure 1**, O1, O3, and P3, which differ in their oxygen stacking sequences: ABAB, ABCABC, and ABBCCA, respectively. Targeting high-capacity NIB cathodes, we limit the complex Na/Vac and TM ordering problem to these three stacking sequences. During cycling, O3 cathodes belong to a glide-connected structural family that includes P3 and O1 phases. O3 phases convert to P3 by a simple translation of the $TMO_2$ slabs by (1/3, 2/3, 0) without breaking any TM-O bonds[22,43]. O1 sits at the other end of the family, since it becomes competitive only when the Na layer is nearly empty at the end of charge[23,27]. We do not consider P2-type phases because the transformation from O3 to P2 requires breaking TM-O bonds and occurs only at synthesis temperatures. In this sense, P2 phases are not electrochemically connected to the O3 family[13,22]. Hybrid stacks of O- and P-type layers, like P2-O3 intergrowths or staged O1-O3 phases, are not considered here because their long repeating units are too complex for systematic Na/Vac order sampling with current computational capabilities.

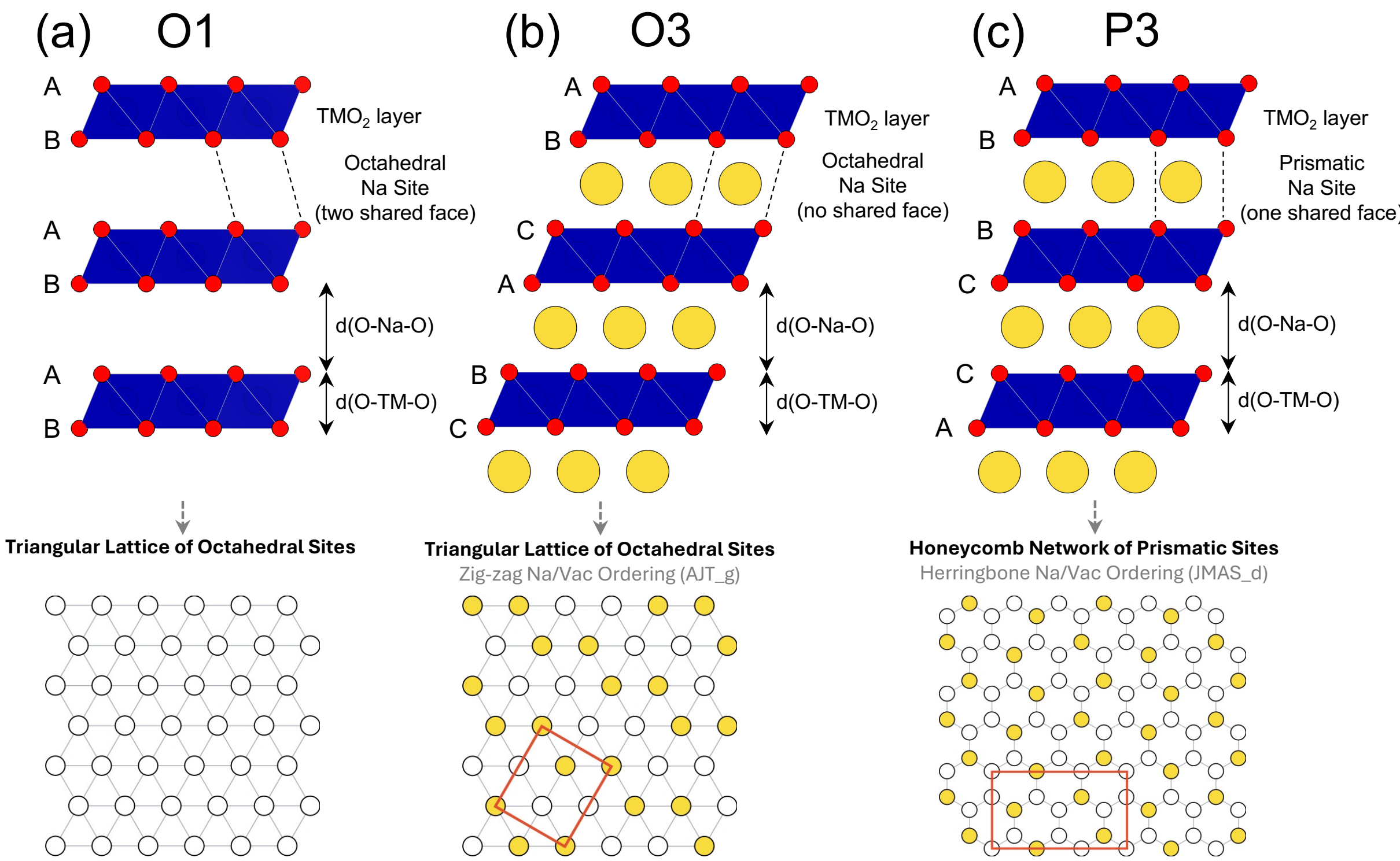

**Figure 1. Three-layer stacking models of $Na_xTMO_2$ in the Delmas notation**[22]: (a) O1, (b) O3, and (c) P3, with the oxygen stacking sequence labeled on each panel. Blue: $TMO_6$ octahedra forming $TMO_2$ slabs; red: O; yellow: Na; arrows: interslab distance d(O-Na-O) and d(O-TM-O). Bottom row: in-plane Na site networks. The orderings in (b) and (c) are one example each of the many possible Na/Vac orderings (filled/open circles: occupied/vacant sites; red outline: unit cell; all orderings in **Figures S1** and **S2**). The O1 phase is shown as an empty network, as it appears only near full desodiation.

This work integrates Na/Vac orderings from prior studies into a single library and calculates them using consistent computational settings across the Na layered oxide family. From the energetics of each system, we construct the phase diagram at 0 K, also known as the convex hull, which identifies stable structures that may feature as distinct steps in voltage profiles. These phase diagrams identify the enthalpically stabilized orderings, the preferred stacking sequence (O1, O3, or P3), and the corresponding equilibrium voltage profile. They distinguish stable ground-state phases from entropy-stabilized phases that may also be observed experimentally. On this basis, we build a uniform map of the O1/O3/P3 family for five single-TM and all ten equimolar binary $Na_xTM_{0.5}TM'_{0.5}O_2$ systems (TM, TM' = Ti, Mn, Fe, Co, Ni), which are an attractive choice from a commercial perspective, and comprehensively evaluate voltages and stable Na/Vac orderings, oxidation states and thermodynamic stability, Jahn–Teller distortions, an energy-density-versus-strain design map, and physical and chemical descriptors of the stacking.

Our computed voltage profiles explained the experimentally observed average voltages for specific $Na_xTM_{0.5}TM'_{0.5}O_2$ across the much larger explored compositional landscape. The computed data further clarify each composition's tendency to form P3 phases, its tendency toward specific transition-metal charge ordering, and the extent to which TM mixing suppresses undesired Jahn–Teller distortion—either by charge transfer that stabilizes Mn or Ni in Jahn–Teller-inactive oxidation states or by dilution with Jahn–Teller-inactive cations. We also show that empirical scalar descriptors, such as the cationic potential and the interslab distance, may be inadequate when Na/Vac ordering governs the O3/P3 competition, indicating that more complex models are needed to anticipate these phase transitions.

## Results

We performed sodium-content-dependent energy calculations for single-TM and bimetallic $TM_{0.5}TM'_{0.5}$ (TM, TM' = Ti, Mn, Fe, Co, Ni) systems by enumerating transition-metal orderings and sodium-vacancy orderings. We examine how voltage curves, phase transitions, lattice parameters, and volume evolve in layered sodium oxides, then relate these trends to sodium–vacancy ordering, transition-metal redox, and electronic-structure effects such as Jahn–Teller distortions.

Energetics and Structural Evolution of Sodium Transition-Metal Oxides on Desodiation

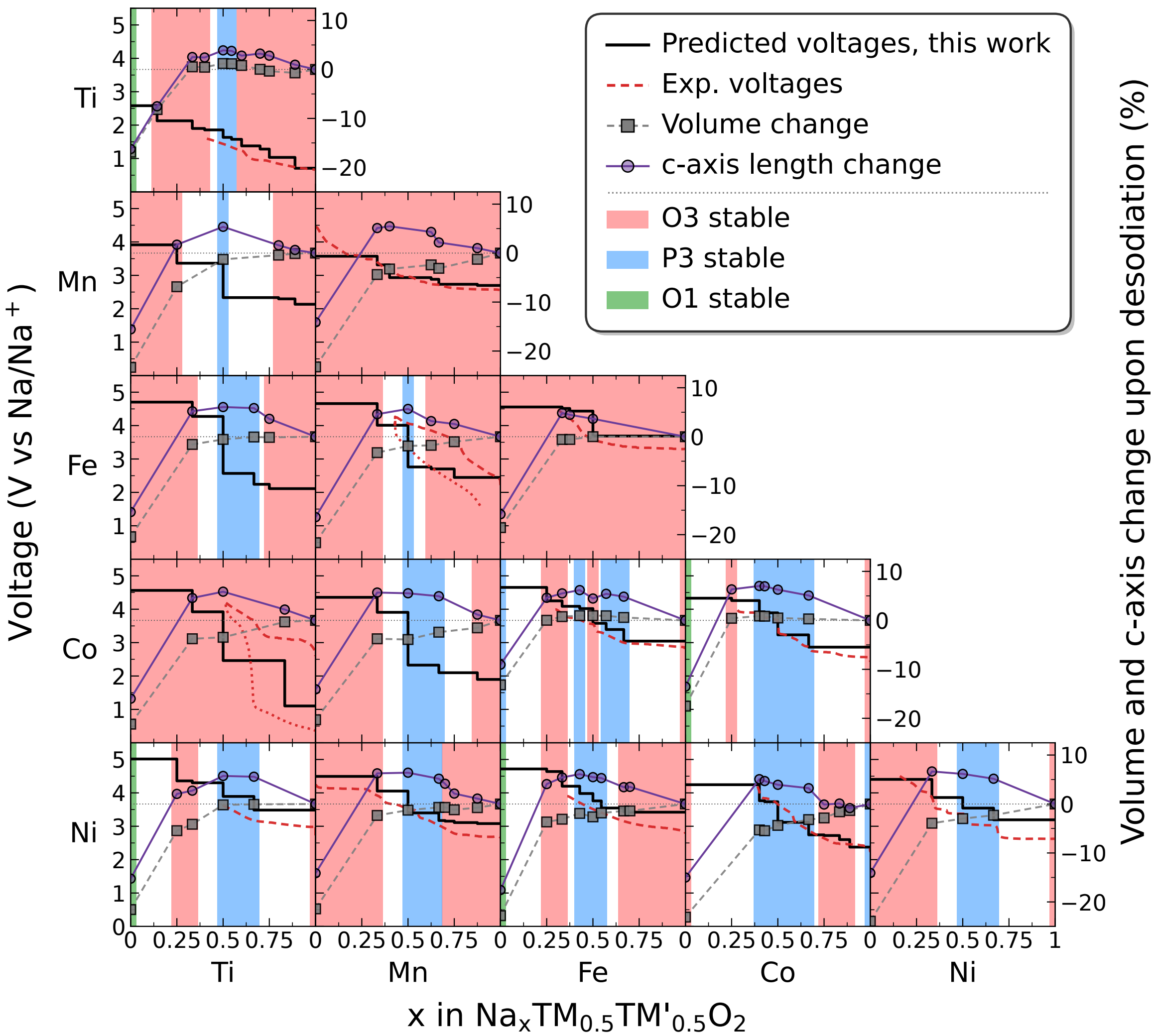

**Figure 2. Predicted voltage profiles and volume and *c*-axis length changes of single- and mixed-transition-metal $Na_xTMO_2$ cathodes**, with single-metal $Na_xTMO_2$ arranged along the diagonal, and equimolar $Na_xTM_{0.5}TM'_{0.5}O_2$; TM = Ti, Mn, Fe, Co, Ni as off-diagonal elements. Black lines display 0 K DFT-predicted voltage (left axis); gray dashed lines show the cell volume per $Na_xTMO_2$ formula unit (right axis); purple lines with circles show the c-axis length per $TMO_2$ layer, *i.e.,* the stacking period divided by the number of $TMO_2$ layers in the cell (right axis); red dashed lines illustrate the experimental galvanostatic profiles where available[13,27,29,41,42,44–48]; where two experimental curves are shown (dashed: charge, dotted: discharge), both are plotted because of the large hysteresis reported in the experiment. Background shading illustrates thermodynamically stable stacking (O1 green, O3 red, P3 blue).

**Figure 2** shows the predicted voltages, together with the variations in volume and *c*-axis length, extracted from 1,025 independent DFT+*U* calculations.

**(i)** **Voltage profiles:** In the predicted voltage profiles, each step indicates a stable Na/Vac ordering, and each plateau represents a two-phase region between neighboring orderings. The voltage difference between steps measures the relative stability of adjacent Na/Vac orderings [21,23,38]. The large, sharp steps in **Figure 2** mark stable Na–vacancy orderings, which likely persist at room temperature, as evidenced by corresponding sharp steps in the experimental voltage profiles, while smaller steps in the computational voltages smear into sloping regions. The predicted average voltages match the measured values for each synthesized compound to within ~0.5 V vs. $Na/Na^+$, the accuracy generally quoted for DFT+*U* intercalation voltages[49]. The predicted average open-circuit voltage of the single-TM Na layered oxides increases in the order Ti < Mn < Co < Ni < Fe, with values of 1.67, 3.10, 3.61, 3.77, and 4.11 V vs. $Na/Na^+$, respectively. Notably, Fe gives the highest average voltage, likely due to charge transfer from the oxygen ligands. Specifically, oxidizing high-spin $d^5$ $Fe^{3+}$ costs the exchange stabilization of its half-filled shell, which pushes the $Fe^{3+/4+}$ level down to the O *2p* band edge, and the hole forms on O-hybridized states[47,50].

**(ii)** **Phase stabilities:** In **Figure 2**, the most stable stacking sequence for each Na–vacancy ordering is color-coded by phase: O3 (red), P3 (blue), and O1 (green). For $x$ = 1, we find the O3 state to be stable across all compositions except $Na_1Co_{0.5}Ni_{0.5}O_2$. The O3→P3 transition differs between single- and mixed-TM systems. Among single-TM oxides, Co, Ni, and Ti favor P3 over O3 at intermediate Na contents ($0.40 \leq x \leq 0.67$, $0.50 \leq x \leq 0.67$, and only $0.50 \leq x \leq 0.55$, respectively), whereas Mn and Fe remain O3. In mixed-TM systems, P3 phases appear stable in most systems, even when Fe is mixed with Mn, consistent with previous experimental reports[13,46]. The only mixed-TM system in which P3 ordering appears unstable is $Na_xTi_{0.5}Co_{0.5}O_2$, where the lowest P3 orderings lie within 2 meV/f.u. of the hull (see **Discussion**).

(iii) **Lattice parameters and volume variations:** The *c*-axis length is given per $TMO_2$ layer, that is, the stacking-period length divided by the number of $TMO_2$ layers it contains (three for O3 and P3, one for O1), so the three stackings can be compared on the same footing. Predicted phase diagrams are in **Figures S3** and **S4** of the Supporting Information, and the simulated X-ray diffractograms of all stable Na/Vac orderings are in **Figures S6** and **S7**. Monitoring volume and c-axis changes (gray dashed and purple lines in **Figure 2**) as sodium content decreases is important because these lattice changes determine the strain the electrode must withstand during cycling. The volume shrinks as Na is extracted, but the Na layer's interlayer distance varies with desodiation level. In all systems, before half of the Na content is extracted, the interlayer distance, governed by the *c* length, mostly increases because decreasing Na content reduces the electrostatic screening between TMO layers. During sodium extraction, the in-plane axes, *a* and *b*, contract more as TMs oxidize, causing a slight volume decrease. Complete Na extraction collapses the interlayer distance because no Na remains in the layer.

We contrast the predictive power of our model with existing experimental data for single-TM layered oxides across varying Na compositions. Similar to early predictions for $Na_xCoO_2$ [23], our data reproduce the relative stability of O3/P3 phases observed in the experiments by Kubota *et al.*[45] and predictions by Kaufman *et al.*[23], except for a small O1-O3 stage not considered here. Kaufman *et al.* identified a hierarchy of closely spaced Na–vacancy orderings (JLK_$\zeta$, JLK_$\eta$, JLK_$\delta$ in **Figures S2** and **S3d**) that form a devil's-staircase-like sequence of ground states, whose antiphase boundaries produce solid-solution-like behavior across the P3 region[23,51]. However, our results partly overturn the picture proposed by Kaufman. The zigzag P3 $Na_xCoO_2$ ordering (termed JMAS_d, see **Figure S2** and **Figure S3d** in SI), based on the experimentally resolved Na/Vac ordering seen in $Na_xNiO_2$ by Steele et al.[29], appears more stable than JLK_$\delta$ by 16 meV per formula unit (meV/f.u.) and replaces it on the phase diagram, while JLK_η at x = 4/7 lies 33 meV/f.u. above the stability line (**Figure S3d**).

For $Na_xNiO_2$, the predicted 0 K phase diagram of **Figure S3e** recovers every experimentally reported ordering except for JMAS_b (**Figure S2**), which lies only 11 meV/f.u. above the stability line, and the calculated voltage curve closely follows the experimentally measured one[29,37]. Consistently, the measured voltage step at $x$ = 2/5, where JMAS_b forms, is only ~0.1 V, compared with ~0.3 V for JMAS_c at $x$ = 1/2 and JMAS_d at $x$ = 2/3 (**Figure 2**).

In $Na_xFeO_2$, the O3 phase persists over the whole composition range we computed and over the range accessible in experiment. Experimentally, reversible cycling is limited to $x$ > 0.5 (a charge cutoff of ~3.5 V vs. $Na/Na^+$). This limit comes from $Fe^{4+}$ migrating into the Na layer on desodiation, not from the thermodynamic instability of the O3 host, which is why our calculations keep O3 stable down to $x$ = 0[13,52].

Chen *et al.* reported three Na–vacancy orderings in $Na_xMnO_2$ at $x$ = 5/8, 1/2, and 1/3 [27]. Our phase diagram recovers two of them as ground states: the $Na_{5/8}$ superstructure (XL_a, **Figure S3b**), including the experimentally observed displacement of Na off its ideal site [26,27], and the $x$ = 1/3 ordering (XC_a, a monoclinic distortion of AJT_d with the same Na/Vac ordering, labeled AJT_d in **Figure S3b**). The $x$ = 1/2 ordering (XC_b, **Figure S3b**) also features partially displaced Na but lies ~45 meV/f.u. (12.9 meV/atom) above the phase-diagram stability line, suggesting this phase is entropy-stabilized during high-temperature synthesis. Below $x$ = 1/3, we predict O3 to remain stable, whereas Chen *et al.* observed an O1 phase with residual Na ($x$ = 1/18), a composition that our largest supercell cannot capture. Apart from these two discrepancies, all predicted ground states are O3-type ($x$ = 0, 1/3, 2/5, 5/8, 2/3, 7/8, and 1), consistent with the O3-derived phases reported across the full composition range [27].

Finally, experiments show that $Na_xTiO_2$ adopts the O3 phase at low voltage (~1 V, $x \geq 0.5$)[44], whereas our calculations predict a narrow P3 window at $0.50 \leq x \leq 0.55$.

We now turn to mixed TM layered oxides and compare our predictions with e*x situ* or *in situ*/*operando* experiments that tracked O3, P3, and O1 phase evolution during cycling. We begin with the Ni-containing materials, which are important for their high operating voltages (>4.3 V vs. $Na/Na^+$; **Figure 2**) [41,42,53,54]. For $Na_xMn_{0.5}Ni_{0.5}O_2$, Komaba *et al.* proposed that the as-synthesized O3 phase transforms on charge via O3 → O′3 → P3 → P′3 → P3″, driven by the $Ni^{2+}/Ni^{4+}$ redox couple with $Mn^{4+}$ inactive[41]. Later *operando* X-ray diffraction (XRD) by Kubota *et al.* instead resolved O3 → O′3 → P′3/P3 → O′3 → O3, with a return to O-type stacking at low Na content[53]. Our predictions reproduce this sequence (**Figures 2** and **S4g**): O3 is stable for $x \geq 0.70$, P3 for $0.50 \leq x \leq 0.67$, and O3 again below $x = 0.33$.

Kubota *et al.* observed that P3 transforms into monoclinic O′3 at $x \approx 0.37$ (interslab distance ~5.67 Å), and that the nearly Na-free O3 phase at the end of charge has a much smaller interslab distance of approximately 4.42 Å[53]. In contrast, the P3″ phase that Komaba *et al.* observed at 4.5 V vs. $Na/Na^+$ by *ex situ* XRD had an anomalously large interslab distance (~7.0 Å), which the authors attributed to solvent co-intercalation rather than a bulk oxide phase[41]. Therefore, this phase is not included in our enumeration effort.

For $Na_xCo_{0.5}Ni_{0.5}O_2$, *in situ* bench-quality XRD (Mo Kα) by Vassilaras *et al.* resolved several O3/P3 two-phase regions on charge[42]. The pristine O3 phase distorts to a monoclinic O′3 structure at $x \approx 0.96$ and retains O-type stacking down to $x \approx 0.84$. P-type stacking then represents a broad range of sodium compositions $0.43 \leq x \leq 0.72$, evolving from monoclinic P′3 (0.66–0.72) to hexagonal P3 (0.58–0.66) and to distorted P″3 (0.43–0.58). At $x \approx 0.41$ (~3.8 V vs. $Na/Na^+$), a P″3–O″3 two-phase reaction restores O3-type stacking, which persists to the end of charge ($x \approx 0.36$). Our phase diagram reproduces this O3 → P3 → O3 sequence, with O3 stable for $0.75 \leq x \leq 0.89$, P3 for $0.40 \leq x \leq 0.67$, and O3 at $x = 0$. The one exception is $x = 1$, where we favor P3 over the lowest-energy O3 ordering by 25 meV/f.u., whereas the as-synthesized material is O3. However, this composition is accessed only through synthesis, since re-sodiation to 2 V vs. $Na/Na^+$ stops near $x \approx 0.85$ in O′3, which we also predict to be O3[42]. Therefore, we suggest that the O3 phase forms during high-temperature synthesis, when slab-glide transformations are thermally active, and entropy can favor O3. The O3 structure is then retained at room temperature, where glide is kinetically frozen, and the small 0 K preference for P3 cannot be realized.

Among the Fe-containing oxides, $Na_xFe_{0.5}Co_{0.5}O_2$ and $Na_xFe_{0.5}Ni_{0.5}O_2$ have the most complete structural records. For $Na_xFe_{0.5}Co_{0.5}O_2$, Kubota *et al.* combined *operando* XRD, *ex situ* synchrotron XRD, and neutron diffraction[45]. During charge, they observed O3 for $x \geq 0.87$, an O3 + P3 two-phase region down to $x \approx 0.64$, and P3 that distorts to P′3 near $x \approx 0.56$ and forms a Na–vacancy ordering at $x = 1/2$, marked by a voltage jump. O3-type stacking returns for $0.30 \leq x \leq 0.39$, after which Fe migration ends the reversible window[45]. Our phase diagram largely reproduces this sequence: O3 at $x = 1$, P3 for $0.57 \leq x \leq 0.67$, O3 at $x = 0.5$, P3 at $x = 0.43$, O3 for $0.25 \leq x \leq 0.33$, and P3 at $x = 0$, where P3, O3, and O1 lie within ~17 meV/f.u. of one another. The main discrepancy is at $x = 0.5$: experiment retains the ordered P′3 phase, whereas we predict the ordering in O3, with the lowest P3 variant only ~16 meV/f.u. (4.6 meV/atom) higher in energy.

For $Na_xFe_{0.5}Ni_{0.5}O_2$, *in situ* XRD by Vassilaras *et al.* shows a direct O3 → P3 transition with no monoclinic distortion: an O3 + P3 two-phase region spans a Na composition region $0.62 \leq x \leq 0.76$, and single-phase P3 persists to the end of the reversible window at $x \approx 0.34$[42]. Our phase diagram (**Figure S4i**) reproduces this sequence, with O3 stable for $x \geq 0.67$ and P3 for $0.43 \leq x \leq 0.55$. Below this, we predict a return to O3 ($0.25 \leq x \leq 0.33$) and O1 at $x = 0$, but both lie beyond the reversible window, which closes at 3.9 V vs. $Na/Na^+$ when $Fe^{4+}$ migrates into the Na layer[42].

In both Fe-containing mixed oxides, the reversible window extends well past the $x \approx 0.5$ limit of $Na_xFeO_2$, to $x \approx 0.30$ for Fe–Co and $x \approx 0.34$ for Fe–Ni, increasing their practical capacities[13,42,45,52]. The stacking sequences, lattice strain, and Na–vacancy orderings discussed so far cannot explain this extension because they do not indicate which transition metal oxidizes first, and hence when the $Fe^{4+}$ that triggers migration appears. In the next section, we extract the oxidation state of each transition metal from the computed magnetic moments and track its evolution during desodiation.

### Oxidation states and charge ordering of the transition metals

Beyond revealing the order in which the transition metals are oxidized as Na extraction progresses, the oxidation states help explain the phase transitions, which depend on the local charge environment established by the Na–vacancy arrangement and the transition metals. We assign each transition-metal ion the oxidation state whose typical magnetic moment best matches its computed value: 3.9 $\mu_B$ for $Mn^{3+}$ and 3.1 $\mu_B$ for $Mn^{4+}$, 4.3 $\mu_B$ for $Fe^{3+}$ and 3.9 $\mu_B$ for $Fe^{4+}$, 0.0 $\mu_B$ for low-spin $Co^{3+}$ and 1.1 $\mu_B$ for $Co^{4+}$, and 1.8, 1.2, and 0.2 $\mu_B$ for $Ni^{2+}$, $Ni^{3+}$, and $Ni^{4+}$, while $Ti^{4+}$ carries no magnetic moment. These reference values are the mean computed moments of the ions in the hull-stable structures of single-TM oxides. They differ from the spin-only values of the formal *d* counts because the on-site moment is integrated inside the projector augmented-wave sphere of the metal atom, and the covalent TM–O bonds delocalize part of the spin density onto the oxygen ligands. We use the moments rather than other less uniquely defined charges, such as Bader charges, because the moment changes by 0.4–1.1 $\mu_B$ per oxidation step, whereas the Bader charge of a TM ion changes by only 0.1–0.3 *e*, as the same covalency keeps it far from the formal charge[45,55]. The moment also separates spin states, which a charge cannot. The same moment-based assignment has been used in previous DFT studies of Na layered oxides and polyanionic positive-electrode materials[4,27,42].

**Figure 3** reports the resulting oxidation states, averaged over all ions of each element in the cell, at selected Na contents during desodiation. Fractional values indicate that two or more oxidation states of the same element coexist, and "n/a" marks compositions where the assigned states do not satisfy charge neutrality. In **Figure 3**, cell colors show the energy above the stability line (convex hull) relative to competing phases reported in the Materials Project database, and asterisks mark compositions reached electrochemically, allowing direct comparison of thermodynamic stability and electrochemical accessibility. Compositions lying well above the convex hull are metastable or unstable and may decompose into competing phases. However, as the asterisks in **Figure 3** show, many of these phases are still observed in electrochemical experiments, because desodiation is topotactic and decomposition is kinetically impeded at room temperature. Entropy contributions are not included and may further stabilize partially disordered phases[23,37,56].

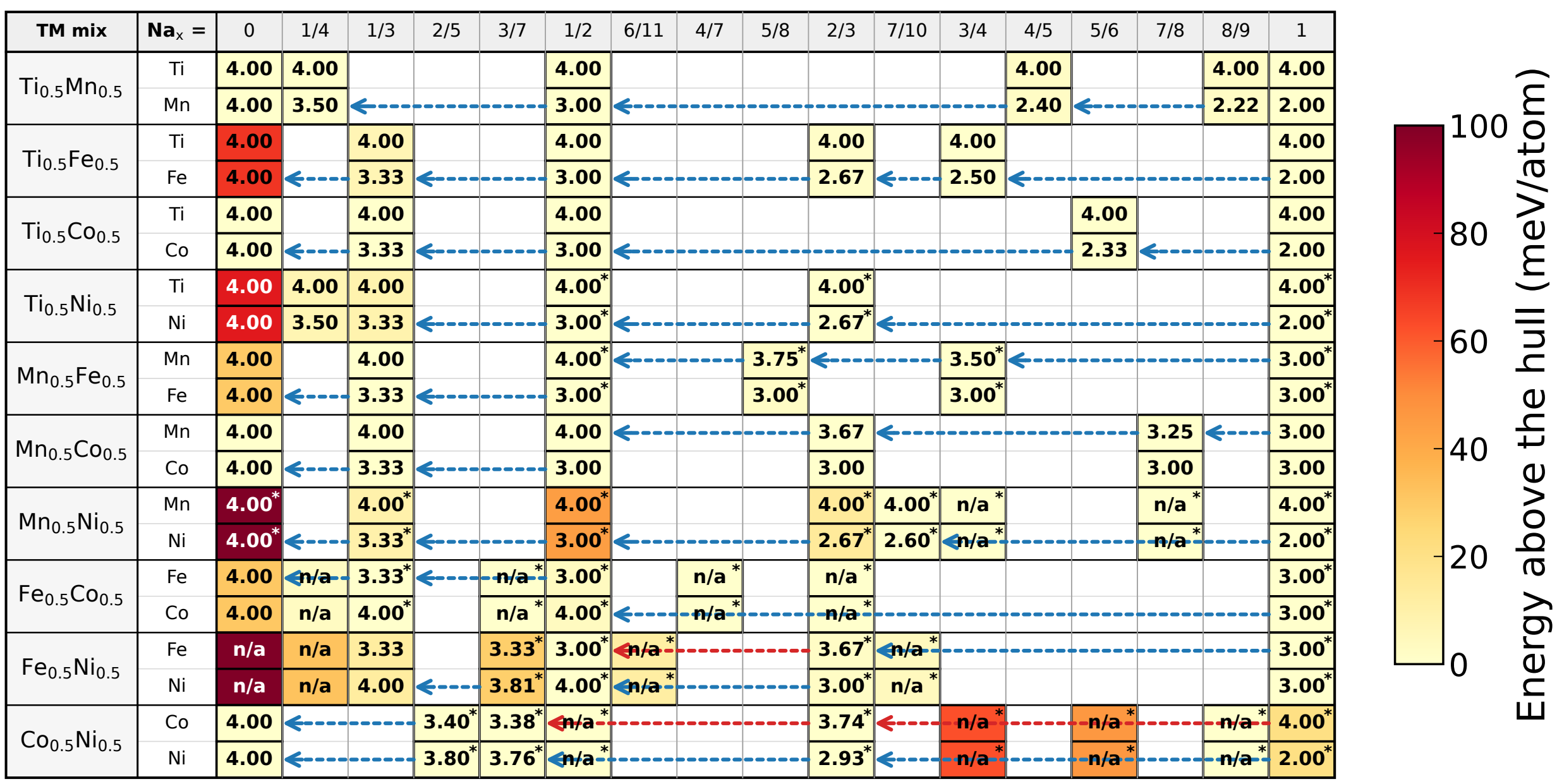

| TM mix | $Na_x$ = | 0 | 1/4 | 1/3 | 2/5 | 3/7 | 1/2 | 6/11 | 4/7 | 5/8 | 2/3 | 7/10 | 3/4 | 4/5 | 5/6 | 7/8 | 8/9 | 1 |
|---|---|---|---|---|---|---|---|---|---|---|---|---|---|---|---|---|---|---|
| $Ti_{0.5}Mn_{0.5}$ | Ti | 4.00 | 4.00 | | | | 4.00 | | | | | | | 4.00 | | | 4.00 | 4.00 |
| | Mn | 4.00 | 3.50 | | | | 3.00 | | | | | | | 2.40 | | | 2.22 | 2.00 |
| $Ti_{0.5}Fe_{0.5}$ | Ti | 4.00 | | 4.00 | | | 4.00 | | | | 4.00 | | 4.00 | | | | | 4.00 |
| | Fe | 4.00 | | 3.33 | | | 3.00 | | | | 2.67 | | 2.50 | | | | | 2.00 |
| $Ti_{0.5}Co_{0.5}$ | Ti | 4.00 | | 4.00 | | | 4.00 | | | | | | | | 4.00 | | | 4.00 |
| | Co | 4.00 | | 3.33 | | | 3.00 | | | | | | | | 2.33 | | | 2.00 |
| $Ti_{0.5}Ni_{0.5}$ | Ti | 4.00 | 4.00 | 4.00 | | | 4.00* | | | | 4.00* | | | | | | | 4.00* |
| | Ni | 4.00 | 3.50 | 3.33 | | | 3.00* | | | | 2.67* | | | | | | | 2.00* |
| $Mn_{0.5}Fe_{0.5}$ | Mn | 4.00 | | 4.00 | | | 4.00* | | | 3.75* | | | 3.50* | | | | | 3.00* |
| | Fe | 4.00 | | 3.33 | | | 3.00* | | | 3.00* | | | 3.00* | | | | | 3.00* |
| $Mn_{0.5}Co_{0.5}$ | Mn | 4.00 | | 4.00 | | | 4.00 | | | | 3.67 | | | | | 3.25 | | 3.00 |
| | Co | 4.00 | | 3.33 | | | 3.00 | | | | 3.00 | | | | | 3.00 | | 3.00 |
| $Mn_{0.5}Ni_{0.5}$ | Mn | 4.00* | | 4.00* | | | 4.00* | | | | 4.00* | 4.00* | n/a* | | | n/a* | | 4.00* |
| | Ni | 4.00* | | 3.33* | | | 3.00* | | | | 2.67* | 2.60* | n/a* | | | n/a* | | 2.00* |
| $Fe_{0.5}Co_{0.5}$ | Fe | 4.00 | n/a | 3.33* | | n/a* | 3.00* | | n/a* | | n/a* | | | | | | | 3.00* |
| | Co | 4.00 | n/a | 4.00* | | n/a* | 4.00* | | n/a* | | n/a* | | | | | | | 3.00* |
| $Fe_{0.5}Ni_{0.5}$ | Fe | n/a | n/a | 3.33 | | 3.33* | 3.00* | n/a* | | | 3.67* | n/a* | | | | | | 3.00* |
| | Ni | n/a | n/a | 4.00 | | 3.81* | 4.00* | n/a* | | | 3.00* | n/a* | | | | | | 3.00* |
| $Co_{0.5}Ni_{0.5}$ | Co | 4.00 | | | 3.40* | 3.38* | n/a* | | | | 3.74* | | n/a* | | n/a* | | n/a* | 4.00* |
| | Ni | 4.00 | | | 3.80* | 3.76* | n/a* | | | | 2.93* | | n/a* | | n/a* | | n/a* | 2.00* |

**Figure 3.** Predicted average TM oxidation states and thermodynamic stability of mixed $Na_x(TM)_{0.5}(TM')_{0.5}O_2$ systems vs. Na content x (two rows per pair: TM top, TM' bottom). Colors show the energy above the convex hull (meV/atom) of each composition relative to the phases in the Materials Project database. Numbers give the average oxidation state of each transition metal, inferred from the DFT on-site magnetic moments; "n/a" marks compositions where the assigned oxidation states violate charge neutrality. Arrows mark redox-active elements during desodiation: blue for oxidation, red for reduction. Asterisks mark stable compositions within the Na range reached by electrochemical (de)sodiation in experiments[41,42,45–47,53].

In **Figure 3**, we first consider the fully sodiated compositions ($x$ = 1). In the single-TM oxides, every transition metal is trivalent, but in the mixed-TM oxides one metal can transfer an electron to the other. In all four Ti-containing oxides, Ti is oxidized to $Ti^{4+}$, and its partner is reduced to $Mn^{2+}$, $Fe^{2+}$, $Co^{2+}$, or $Ni^{2+}$. Likewise, Ni is reduced by Mn and Co, giving $Mn^{4+}/Ni^{2+}$ in $NaMn_{0.5}Ni_{0.5}O_2$ and $Co^{4+}/Ni^{2+}$ in $NaCo_{0.5}Ni_{0.5}O_2$. In the remaining Mn–Fe, Mn–Co, Fe–Co, and Fe–Ni oxides, both metals stay trivalent.

During desodiation, TMs are oxidized sequentially (blue arrows, **Figure 3**). In the Ti-containing oxides and in $Na_xMn_{0.5}Ni_{0.5}O_2$, $Ti^{4+}$ or $Mn^{4+}$ remain inactive, and only the divalent partner is oxidized to 3+ at $x = 0.5$ and to 4+ at $x = 0$, respectively. In oxides where both metals start trivalent, one is fully oxidized to 4+ by $x = 0.5$ before the other TM becomes active, in the order Mn before Fe or Co, Co before Fe, and Ni before Fe. The Ni-containing oxides with Fe and Co are more complex, because the partner TM is partly reduced, while Ni is oxidized (red arrows, **Figure 3**).

In $Na_xFe_{0.5}Ni_{0.5}O_2$, Fe starts as $Fe^{3+}$, rises to +3.67 at $x = 2/3$, and returns to +3 at $x = 1/2$. In $Na_xCo_{0.5}Ni_{0.5}O_2$, Co decreases from +4 at $x = 1$ to about +3.38 at $x = 3/7$, while Ni is oxidized from +2 to nearly +4. We attribute these changes in the partner metal to Ni–O covalency rather than to a separate redox step: the holes created on charge enter O *2p* states hybridized with Ni *3d*, which also shifts the magnetic moments of the neighboring Fe and Co[47,57].

We contrast the moment-derived oxidation states of TMs with electron energy-loss spectroscopy (EELS), Mössbauer, and X-ray absorption spectroscopy (XAS) measurements, which are available for six of the ten mixed oxides (**Table S1**). In general, our assignments agree well with the experiments.

For $Na_xMn_{0.5}Ni_{0.5}O_2$ and $Na_xTi_{0.5}Ni_{0.5}O_2$, experiments confirm the predicted $Ni^{2+}/Mn^{4+}$ and $Ni^{2+}/Ti^{4+}$ pristine states, with only Ni oxidized on charge and the tetravalent partner inactive[41,47]. In $Na_xMn_{0.5}Fe_{0.5}O_2$, $Mn^{3+}$ is oxidized first and $Fe^{3+}$ only near the end of charge[46]. In $Na_xFe_{0.5}Co_{0.5}O_2$, the Co K-edge shifts during the first 45% of Na extraction, whereas the Fe K-edge does not shift until one-third of the Na has been removed[45], consistent with our prediction that Co is oxidized first. Thus, Co substitution delays the formation of $Fe^{4+}$, which explains the extended reversible window discussed in the previous section. For $Na_xFe_{0.5}Ni_{0.5}O_2$, Mössbauer and X-ray absorption spectroscopy show that Fe remains $Fe^{3+}$ while Ni is oxidized[47,57], in agreement with our assignment at $x$ = 1/2, where Ni is fully oxidized to $Ni^{4+}$ and Fe is trivalent. The transient Fe oxidation state of +3.67 at $x$ = 2/3 likely reflects the Ni–O covalency discussed above rather than a separate Fe redox step, since Fe returns to +3 at $x$ = 1/2 while Ni is oxidized further. The only two discrepancies, a uniform $Co^{3+}/Ni^{3+}$ rather than $Co^{4+}/Ni^{2+}$ state in pristine $NaCo_{0.5}Ni_{0.5}O_2$ and Fe rather than Ni oxidation during the first half of charge of $Na_xFe_{0.5}Ni_{0.5}O_2$, come from the *ex situ* EELS study of Vassilaras *et al.*[42] and are examined in the Discussion.

**Figure 3** shows which metal is oxidized at each Na content, but not why metals oxidize in this order or why electrons transfer between metals in some pristine compounds. To address this, **Figure 4a** splits the predicted voltage of each system into averages over the first ($x$ = 1 → 0.5) and second ($x$ = 0.5 → 0) halves of charge, each labeled with its active redox couple from **Figure 3**. Because the open-circuit voltage reflects the energy of the active redox couple relative to $Na/Na^+$ [58], it indicates how far the corresponding TM states lie below the Na Fermi level (**Figure 4b**). A caveat is that this measure also depends on TM–O covalency and electrostatics, which vary with the partner metal and Na content. To minimize this ambiguity, we compare redox couples only within the same half-charge state.

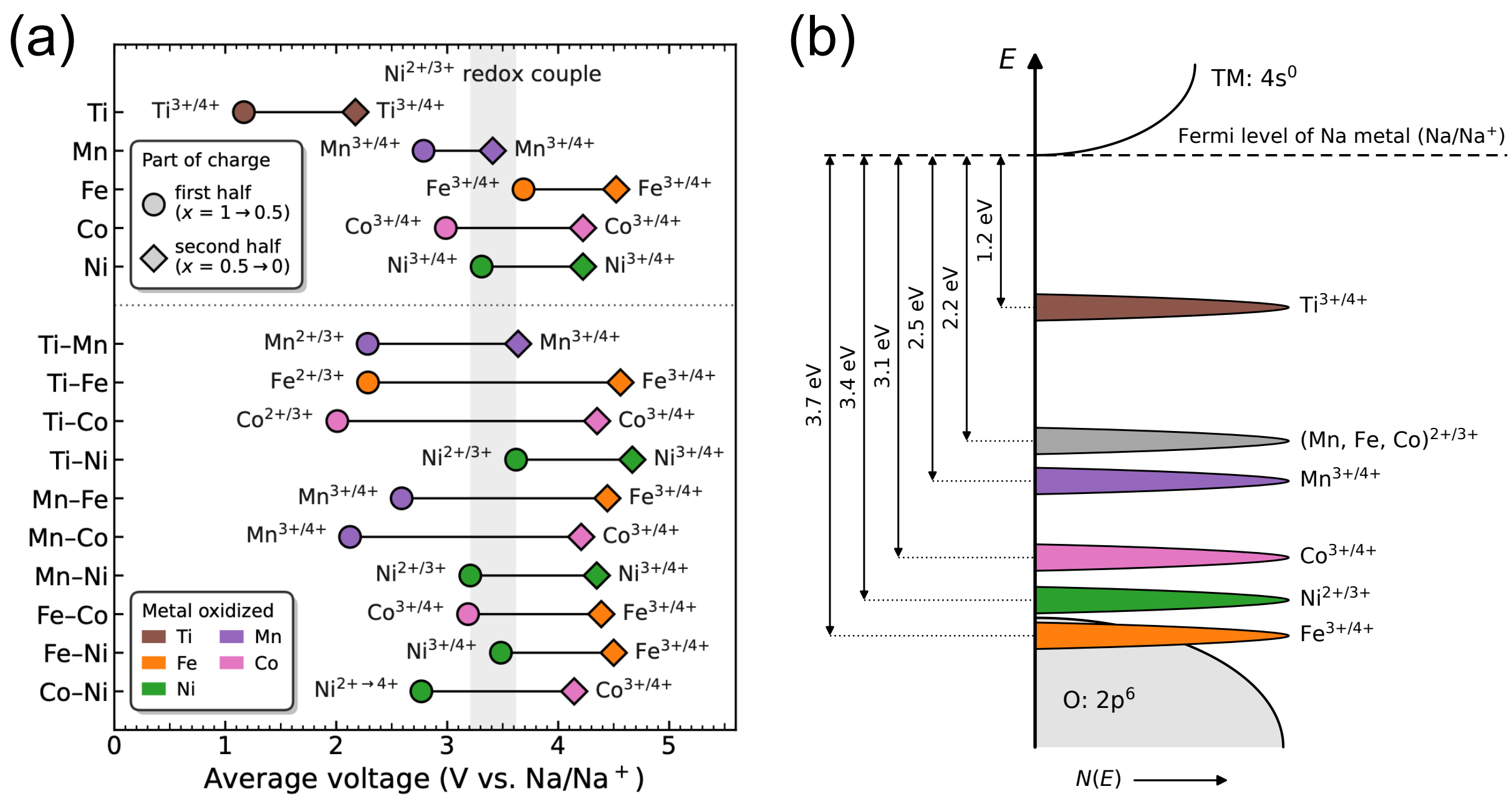

**Figure 4.** (a) Average voltage of each redox couple over the first (circles) and second (diamonds) halves of charge, calculated from the stable compositions in **Figure 2**. Colors identify the oxidized transition metal, labels give the active redox couple, and the gray band marks the voltage range of the $Ni^{2+/3+}$ couple. (b) Schematic energy diagram of the redox couples relative to the Fermi level of Na metal ($Na/Na^+$). Only the depths of the couples are to scale, set by their average first-half voltages from panel (a). *N(E)* density of states.

Over the first half of charge, the redox couples in **Figure 4a** follow the order $Ti^{3+/4+}$ (1.2 V vs. $Na/Na^+$) < $Mn^{3+/4+}$ (2.1–2.8 V) < $Co^{3+/4+}$ (3.0–3.2 V) < $Ni^{2+/3+}$ (3.2–3.6 V) < $Fe^{3+/4+}$ (3.7 V), with the $M^{2+/3+}$ couples of Mn, Fe, and Co at 2.0–2.3 V. This ordering explains the pristine states in **Figure 3**. An electron transfers from one metal to its partner when the first metal's 3+/4+ couple lies at a lower voltage, that is, higher in energy (**Figure 4b**), than the second metal's 2+/3+ couple, because oxidizing the first metal costs less energy than reducing the second gains. $Ti^{3+/4+}$ lies lowest in voltage vs. $Na/Na^+$, donating an electron to every partner. $Mn^{3+/4+}$ and $Co^{3+/4+}$ both lie below $Ni^{2+/3+}$, so Mn or Co donates to Ni, although for Co the margin is small, making the $Co^{4+}/Ni^{2+}$ assignment in $NaCo_{0.5}Ni_{0.5}O_2$ less certain than the others. In the remaining oxides, no 3+/4+ couple lies at a lower voltage than the partner's 2+/3+ couple, and both metals stay trivalent.

The voltage ordering of the redox couples of **Figure 4a** also explains the redox sequence on charge: redox couples are depleted from low to high voltage, so Mn is oxidized before Co and Fe, Co before Fe, and Ni before Fe, since the $Ni^{3+/4+}$ couple (3.3–3.5 V vs. $Na/Na^+$, see below) also lies at a lower voltage than $Fe^{3+/4+}$. $Fe^{3+/4+}$ is always last because removing an electron from high-spin $d^5$ $Fe^{3+}$ breaks four of its ten parallel-spin pairs, the largest loss of exchange stabilization among the trivalent ions considered here, which pushes the $Fe^{3+/4+}$ level down to the top of the O *2p* band (**Figure 4b**) [50,57,58].

**Figure 4a** also shows that a higher oxidation state does not always give a higher voltage. Over the first half of charge, the $Ni^{3+/4+}$ couples of $Na_xNiO_2$ (3.31 V vs. $Na/Na^+$) and $Na_xFe_{0.5}Ni_{0.5}O_2$ (3.48 V vs. $Na/Na^+$) fall within the $Ni^{2+/3+}$ range (gray band), and the $Ni^{2+/3+}$ couple of $Na_xTi_{0.5}Ni_{0.5}O_2$ (3.62 V vs. $Na/Na^+$) lies even higher. This "redox potential paradox" was reported by Nanba *et al.* [47], who measured a higher average voltage for $Na_xTi_{0.5}Ni_{0.5}O_2$ (3.11 V vs. $Na/Na^+$) than for $Na_xFe_{0.5}Ni_{0.5}O_2$ (2.99 V vs. $Na/Na^+$) over the same Na range and attributed the low $Ni^{3+/4+}$ voltage to the large O *2p* contribution (>80%) to this redox couple. We accurately reproduce this paradox, with a voltage difference of 0.14 V between the two compounds compared with 0.12 V in experiment.

Because each redox couple varies little with the partner metal, **Figure 4** can predict the pristine states and redox sequences of compositions not calculated here. In $NaNi_{1/3}Fe_{1/3}Mn_{1/3}O_2$, for example, $Mn^{3+/4+}$ lies below $Ni^{2+/3+}$, so Mn should donate an electron to Ni, giving a pristine $Ni^{2+}/Fe^{3+}/Mn^{4+}$ state. Since both Ni couples lie below $Fe^{3+/4+}$, Ni should be oxidized first and Fe only near the end of charge, with $Mn^{4+}$ inactive throughout. This is the sequence observed by XAS: Ni redox dominates up to 4.0 V vs. $Na/Na^+$, $Fe^{3+/4+}$ is activated above this voltage, and Mn remains inactive[59,60].

## Local Jahn–Teller distortions of the $TMO_6$ octahedra

TM oxidation states also affect the local structure of the $TMO_6$ octahedra. Ions with a single $e_g$ electron, high-spin $Mn^{3+}$ ($t_{2g}^3e_g^1$) and low-spin $Ni^{3+}$ ($t_{2g}^6e_g^1$), are Jahn–Teller active and elongate two of their six TM–O bonds. High-spin $Fe^{4+}$ ($t_{2g}^3e_g^1$) is formally Jahn–Teller active as well, but its $TMO_6$ octahedra distort less because one hole is hybridized with oxygen atoms, whereas $Mn^{4+}$, $Ni^{2+}$, $Ni^{4+}$, low-spin $Co^{3+}$, $Fe^{3+}$, and $Ti^{4+}$ retain regular octahedra. When the elongation axes align within the TM layer, the distortion becomes cooperative, lowering the lattice symmetry and introducing anisotropic strain, as in the monoclinic O′3 $NaNiO_2$ phase[61]. Thus, the redox sequence of **Figure 3** determines where and when Jahn–Teller distortions can occur, which we quantify next for each octahedron in the stable structures (**Figure 5**).

To quantify the distortion of each octahedron, we use the Van Vleck modes $Q_2$ and $Q_3$, derived from the lengths of the three O–TM–O axes[62]. $Q_3$ is the tetragonal mode, in which one axis elongates and the other two shorten, and $Q_2$ is the orthorhombic mode, in which one axis elongates, a second shortens, and the third is unchanged. The bond-length distortion of each octahedron can thus be mapped onto the ($Q_2$, $Q_3$) plane in **Figure 5**, where the radius $\rho = (Q_2^2+Q_3^2)^{1/2}$ gives the distortion magnitude, and the polar angle $\varphi = \arctan(Q_2/Q_3)$ gives its type. Octahedra elongated along one axis lie at $\varphi$ = 0°, 120°, or 240° (dotted lines), and those compressed along one axis lie at $\varphi$ = 60°, 180°, or 300°. Because the same reference axes are used for all octahedra in a structure, octahedra elongated along the same direction appear at the same angle. Because $\rho$ is not an intuitive quantity, we added **Figure S5** to complement the Van Vleck analysis with the mean axial-to-equatorial bond-length ratio for each element as a function of Na content.

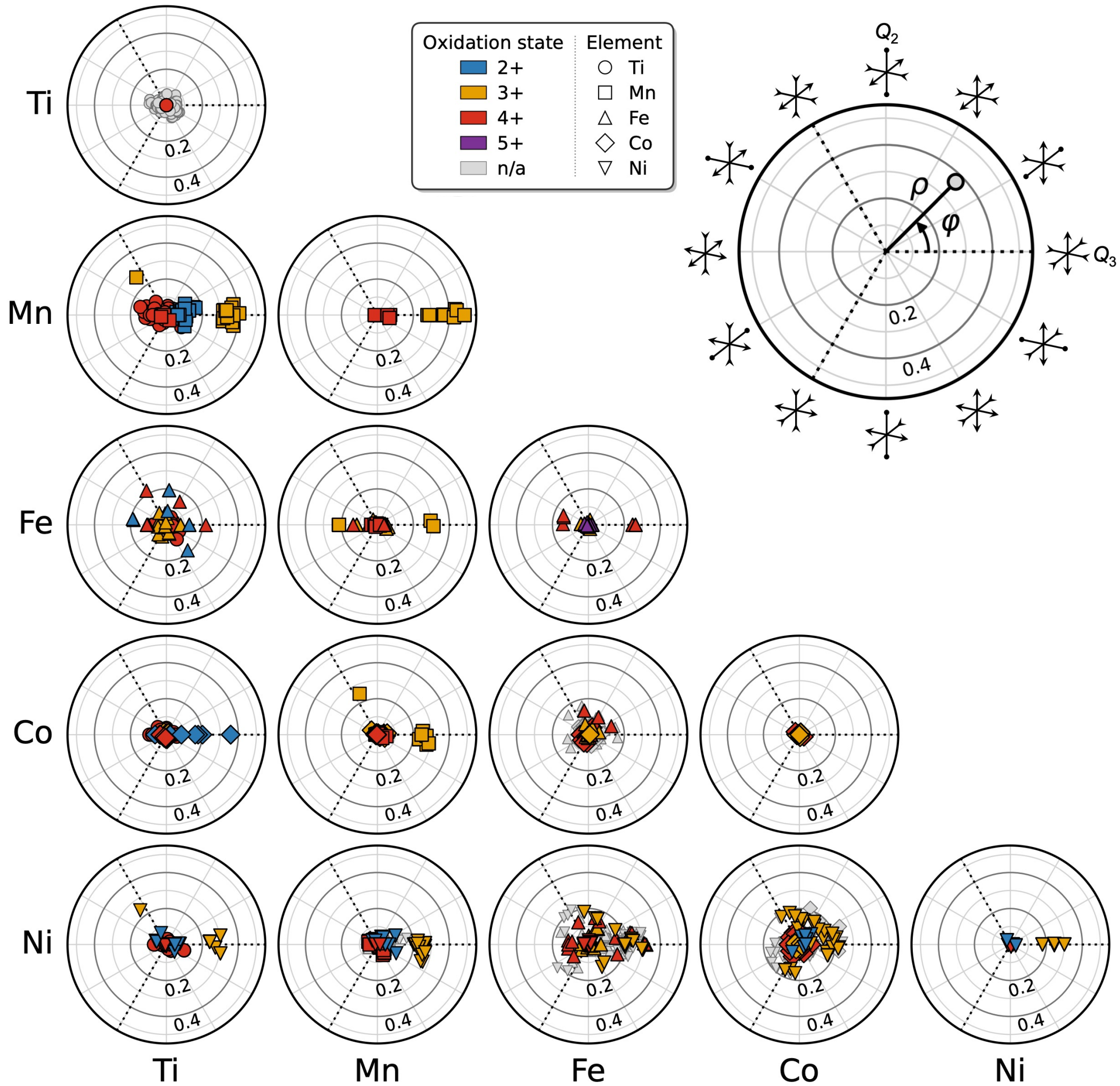

**Figure 5.** $Q_2$–$Q_3$ Van Vleck distortion modes of the $TMO_6$ octahedra in the stable structures of all fifteen systems, arranged as in **Figure 2**. Each point is one octahedron, with radius ρ (Å) giving the distortion magnitude and angle φ its type. The schematic (top right) shows the O–TM–O axes every 30° of φ (outward arrows: elongation; inward: compression; dots: unchanged); dotted lines (φ = 0°, 120°, 240°) mark elongation along each axis. Colors denote oxidation state and markers the element; gray (n/a) marks structures where oxidation states violate charge neutrality.

**Figure 5** shows that the distortion of each octahedron is governed mainly by the oxidation state of its transition metal: only ions with a single $e_g$ electron distort strongly. These are high-spin $d^4$ $Mn^{3+}$ ($t_{2g}^3e_g^1$, yellow squares) and low-spin $d^7$ $Ni^{3+}$ ($t_{2g}^6e_g^1$, yellow inverted triangles), which generally cluster near $\varphi = 0°$ with finite $\rho$, the signature of coherent tetragonal elongation. All 99 $Mn^{3+}$ octahedra are distorted ($0.21 \leq \rho \leq 0.48$ Å), followed by the 101 $Ni^{3+}$ octahedra ($0.06 \leq \rho \leq 0.31$ Å), whereas $Mn^{4+}$, $Ni^{4+}$, $Co^{3+}$, and $Co^{4+}$ remain nearly regular ($\rho < 0.08$ Å). Of the 387 octahedra with $\rho \geq 0.1$ Å, 315 (81%) are elongated along a single axis, with the two long TM–O bonds *trans* to each other, as expected for a Jahn–Teller distortion. Moreover, $\varphi$ deviates from the ideal tetragonal value by only ~2° on average for $Mn^{3+}$ and ~7° for $Ni^{3+}$, confirming nearly pure tetragonal elongation along one axis. The calculated Jahn–Teller distortions also match experiment. For example, the calculated Mn–O bond lengths in $NaMnO_2$ are 1.96 Å (4×) and 2.38 Å (2×), and the calculated Ni–O bond lengths in $NaNiO_2$ are 1.92 Å (4×) and 2.18 Å (2×). The refined Mn–O bond lengths of $NaMnO_2$ (1.94 and 2.39 Å)[63] and the splitting of the Ni–O bond lengths in $NaNiO_2$ (0.24 Å at room temperature)[61] correspond to $\rho$ = 0.52 Å and 0.28 Å, which are close to our values of 0.48 Å and 0.30 Å.

High-spin $d^4$ $Fe^{4+}$ carries the same single $e_g$ electron. In $Na_xFeO_2$, the $Fe^{4+}$ octahedra (red triangle) at x = 0 and x = 0.5 are elongated with $\rho \approx 0.25$ Å, which is comparable to that of $Ni^{3+}$. However, in the mixed-TM oxides, the $Fe^{4+}$ distortion is smaller ($\rho$ = 0.12 Å on average) and shows no preferred type, meaning compressed and elongated octahedra occur in equal numbers. This agrees with the non-cooperative Jahn–Teller distortion of high-spin $Fe^{4+}$, inferred from the large Debye–Waller factor of the Fe–O bonds in the EXAFS analysis of P2-type $Na_xFe_{0.5}Mn_{0.5}O_2$[46]. $Ti^{4+}$ is JT-silent and responds only "elastically" to its distorted neighbors. For example, in $Na_xTi_{0.5}Mn_{0.5}O_2$, $Ti^{4+}$ is compressed by up to $\rho$ = 0.15 Å along the elongation axis of the neighboring $Mn^{3+}$. Co is JT-silent, since its electronic state is mostly low-spin $d^6$ $Co^{3+}$ or $d^5$ $Co^{4+}$, except at the sodiated end of $Na_xTi_{0.5}Co_{0.5}O_2$, where the $Ti^{4+}$ spectator makes $Co^{2+}$ (blue diamond) the analog of $Ni^{3+}$. The computed moments of ≈1 $\mu_B$ at x = 1 indicate low-spin $d^7$ $Co^{2+}$, and its single $e_g$ electron drives a cooperative elongation with $\rho$ = 0.36 Å, larger than the $\rho$ of $Ni^{3+}$ in $NaNiO_2$ (0.30 Å).

Because $\varphi$ identifies the elongation axis of each octahedron, **Figure 5** also reveals how the Jahn–Teller distortions can order. In the stable structures, the elongated axes form rows within the TM layer that are either all parallel "collinear" (*i.e.*, a single $\varphi$) or alternate between two directions, "zig-zag" (two $\varphi$ values)[28]. In $Na_xMnO_2$ and $Na_xNiO_2$, at every Na content, all elongated octahedra in each structure fall on the same dotted line in **Figure 5**, indicating collinear ordering. At intermediate Na composition, this collinear ordering persists, but mixed-valence charge ordering can divide the layer into Jahn–Teller-active and -inactive sites. At $x \leq 1/2$, in $Na_xMnO_2$, Chen *et al.* observed complete charge separation, with $Mn^{3+}$ and $Mn^{4+}$ stripes forming alternating superplanes across successive $MnO_2$ layers[27].

Even for the same ion, the distortion magnitude can vary widely ($\rho$ = 0.21–0.48 Å for $Mn^{3+}$ and 0.06–0.31 Å for $Ni^{3+}$, in **Figure 5**), a spread linked to the collinear ordering described above. The magnitude of the distortion is not a fixed property of the ion, since it increases with the number of neighbors elongated in the same direction (ordering labels as in **Figures S3 and S4**). As the fraction of the JT-active ion increases with x, $\rho$ of $Mn^{3+}$ increases from 0.28 Å at x = 1/3 to 0.48 Å at x = 1 in $Na_xMnO_2$, and $\rho$ of $Ni^{3+}$ increases from 0.18 Å at x = 1/3 to 0.30 Å at x = 1 in $Na_xNiO_2$. For $Mn^{3+}$, $\rho$ is 0.28 Å in $Na_{1/3}MnO_2$ (AJT_d) and 0.30 Å in $Na_{1/4}Ti_{0.5}Mn_{0.5}O_2$ (AJT_b), where none of the six TM neighbors is elongated along the same axis, but 0.43 Å in $Na_{7/8}$MnO2 (AJT_q) and 0.48 Å in $NaMnO_2$ (JMAS_e), where all six neighbors are elongated along the same axis. $Ni^{3+}$ follows the same trend, from 0.18 Å in $Na_{1/3}NiO_2$ and 0.21 Å in $Na_{1/3}Mn_{0.5}Ni_{0.5}O_2$ (both XC_a) to 0.30 Å in $NaNiO_2$ (JMAS_e). Thus, about 40% of the distortion in $NaMnO_2$ and $NaNiO_2$ arises from cooperative alignment with neighboring octahedra, while the remaining ~60% is intrinsic to the individual Jahn–Teller ion.

Mixing transition metals suppresses the Jahn–Teller distortion through two distinct mechanisms. First, charge transfer between the two metals can eliminate the Jahn–Teller-active 3+ state altogether. In $NaMn_{0.5}Ni_{0.5}O_2$ and $NaTi_{0.5}Ni_{0.5}O_2$, for example, the sodiated state adopts $Mn^{4+}/Ni^{2+}$ and $Ti^{4+}/Ni^{2+}$ rather than two trivalent ions, so neither metal is Jahn–Teller active and the strong distortions of the $NaMnO_2$ and $NaNiO_2$ endmembers largely disappear. On charge, Ni passes through the Jahn–Teller-active $Ni^{3+}$ redox state only at intermediate Na contents, where $Ni^{3+}$ octahedra are interleaved with rigid $Mn^{4+}$ (or $Ti^{4+}$) octahedra. Therefore, the distortion is transient and vanishes once Ni is oxidized to Jahn–Teller-inactive $Ni^{4+}$, as shown in **Figure S5** (displaying $\rho$ against Na content). However, charge transfer does not always suppress the distortion. In $NaTi_{0.5}Co_{0.5}O_2$, it produces low-spin $Co^{2+}$ ($t_{2g}^6e_g^1$), which is itself Jahn–Teller active, so the sodiated endmember distorts. This low-spin state is, however, exceptional: at $x$ = 5/6, four of the six $Co^{2+}$ ions are high-spin, and experiment assigns high-spin $Co^{2+}$ to the pristine O3 phase[48]. Second, even when the Jahn–Teller-active ion remains, diluting it with an inactive metal disrupts cooperativity. In $NaMn_{0.5}Fe_{0.5}O_2$ and $NaMn_{0.5}Co_{0.5}O_2$, Mn remains trivalent at high Na content, but Jahn–Teller-inactive $Fe^{3+}$ or $Co^{3+}$ octahedra separate the $Mn^{3+}$ octahedra, reducing the average $\rho$ of $Mn^{3+}$ from 0.48 Å in $NaMnO_2$ to 0.21 and 0.25 Å, respectively.

Furthermore, in $Na_xFe_{0.5}Co_{0.5}O_2$, $Na_xFe_{0.5}Ni_{0.5}O_2$, and $Na_xCo_{0.5}Ni_{0.5}O_2$, the $Ni^{3+}$ (low-spin $d^7$) and $Fe^{4+}$ (high-spin $d^4$) octahedra do not elongate along a common axis: fewer than half of the distorted octahedra ($\rho \geq 0.1$ Å) lie within 10° of a pure tetragonal elongation (dotted lines in **Figure 5**), compared with all of them in $Na_xMnO_2$ and $Na_xNiO_2$. The remaining octahedra are compressed or intermediate between elongation and compression. Therefore, $Ni^{3+}$ and $Fe^{4+}$ show no preferred elongation axis in these mixed oxides, so the cooperative Jahn–Teller distortion is lost and only a local distortion remains (average $\rho$ = 0.13–0.19 Å). This suppression should reduce the anisotropic lattice strain during cycling. For example, pristine $NaNi_{0.5}Mn_{0.5}O_2$ is rhombohedral, and the monoclinic O′3 and P′3 phases that form on charge show in-plane distortions, defined as the mismatch between $a/\sqrt{3}$ and $b$ of the monoclinic cell, of only ~1.5% and ~0.5%, respectively [41]. By comparison, the same distortion, calculated from reported lattice parameters, is ~8% in $NaNiO_2$ and ~15% in $NaMnO_2$[29,63].

## Discussion

Our first-principles results for single-transition-metal Na layered oxides show only minor disagreements with experimental reports, such as the narrow P3 window of $Na_{0.5}TiO_2$[44] and O3 rather than O1 at the deeply desodiated $MnO_2$ endpoint[27]. In both cases, the energy difference is only a few meV/f.u.; at $Na_{0.5}TiO_2$, the stable P3 ordering lies only 4.8 meV/f.u. below the lowest O3 ordering, and at $MnO_2$, O1 lies 8.2 meV/f.u. above O3. Kinetic effects can easily overcome such small energy differences. Note that the O3↔P3 phase transition has been observed to proceed by a nucleation-limited two-phase glide, so the O3 phase can persist during cycling where P3 is the equilibrium phase[23,25,29,51].

For $Na_xMn_{0.5}Fe_{0.5}O_2$, the predicted and measured voltages in **Figure 2** differ appreciably. However, the O3 phase shows a ~1 V hysteresis between charge and discharge [46]. The measured charge curve in **Figure 2** lies ~0.7 V above our prediction, placing the predicted equilibrium voltage between the charge and discharge curves [46]. Given this uncertainty, a direct comparison is difficult; nonetheless, the predicted phase diagrams (**Figures S3** and **S4**) provide the equilibrium reference for identifying metastable pathways in experiments. $Na_xTi_{0.5}Co_{0.5}O_2$ is a similar case. Maletti *et al.* synthesized both its O3 and P3 phases and observed the O3→P3 transformation on desodiation, but with a large voltage hysteresis (**Figure 2**) that they attribute to kinetics[48]. Since the computed O3–P3 energy difference is marginal (within 2 meV/f.u. at $x = 1/2$ and 2/3), this kinetically controlled transformation cannot be compared directly with our equilibrium hull. DFT inaccuracies aside, the predicted phase diagrams describe the 0 K enthalpic ground states. Configurational entropy, accessed by Na/vacancy and TM disorder, and vibrational entropy can shift the balance among closely energy-spaced orderings at finite temperature, and the ordering-resolved energies compiled in this paper provide the reference data needed to quantify these effects in future work.

Where experimental oxidation-state assignments for the TM redox reactions differ from our predictions, the experiments are themselves not fully consistent. Our Co–Ni (in $Na_xCo_{0.5}Ni_{0.5}O_2$) and Fe–Ni (in $Na_xFe_{0.5}Ni_{0.5}O_2$) assignments differ in part from those of Vassilaras *et al.*[42] For Co–Ni, our magnetic moments suggest a disproportionated $Co^{4+}/Ni^{2+}$ discharged state, whereas their DFT and EELS results support a uniform $Co^{3+}/Ni^{3+}$ charge distribution. For Fe–Ni, Vassilaras *et al.* assign the first half of the charge to Fe oxidation, whereas our simulations indicate that Ni carries the redox activity and Fe is trivalent again at $x = 1/2$, after the transient +3.67 at $x = 2/3$ that we attributed to Ni–O covalency in the Results. However, consistent with our prediction for $Na_xFe_{0.5}Ni_{0.5}O_2$, Nanba *et al.* later showed with Mössbauer and X-ray absorption spectroscopies that $Fe^{3+}$ is retained through half desodiation, with the holes entering hybridized Ni–O states[47,57]. We consider this result more reliable, as EELS probes only local regions of cathode particles, whereas Mössbauer and XAS average over the entire bulk. Nonetheless, our result partly agrees with the EELS picture at the beginning of charge, since the transient +3.67 of Fe at $x = 2/3$ indicates that part of the holes reside on the Fe–O bonds before Ni is fully oxidized. For Co–Ni, the open question is whether the sodiated state is uniformly trivalent or charge-transferred to $Co^{4+}/Ni^{2+}$. If the charge transfer fluctuates in time, as recently resolved for the Ni charge order in $Na_{2/3}NiO_2$, time-averaged spectra cannot distinguish the two pictures; settling the question will require local probes sensitive to dynamics, such as NMR, total scattering, or inelastic spectroscopies[37].

By comparing predicted phase stability, voltages, oxidation states, and Jahn–Teller distortions with experiments, we evaluate the fifteen $Na_xTMO_2$ systems as positive electrode materials. A practical positive electrode requires both a high energy density, set by the voltage of the active redox couples (of **Figures 2** and **4**) and the capacity, and a small volumetric strain during cycling, governed by the phase transitions and Jahn–Teller distortions of the host (**Figures 2** and **5**). **Figure 6** captures this trade-off by plotting the two quantities against each other during desodiation (charging), starting from the fully discharged state ($x = 1$), which is assigned zero energy density and zero strain. At each stable sodium composition (**Figures S3** and **S4**), the gravimetric energy density is the integral of the predicted voltage profile from $x = 1$ to that composition, and the peak volumetric strain is the largest volume change relative to $x = 1$ reached up to that composition; both quantities are therefore cumulative and cannot decrease along a curve. The fully desodiated end member ($x$ = 0) is excluded from **Figure 6** for all fifteen systems, as it undergoes a large volume contraction that is impractical for reversible cycling.

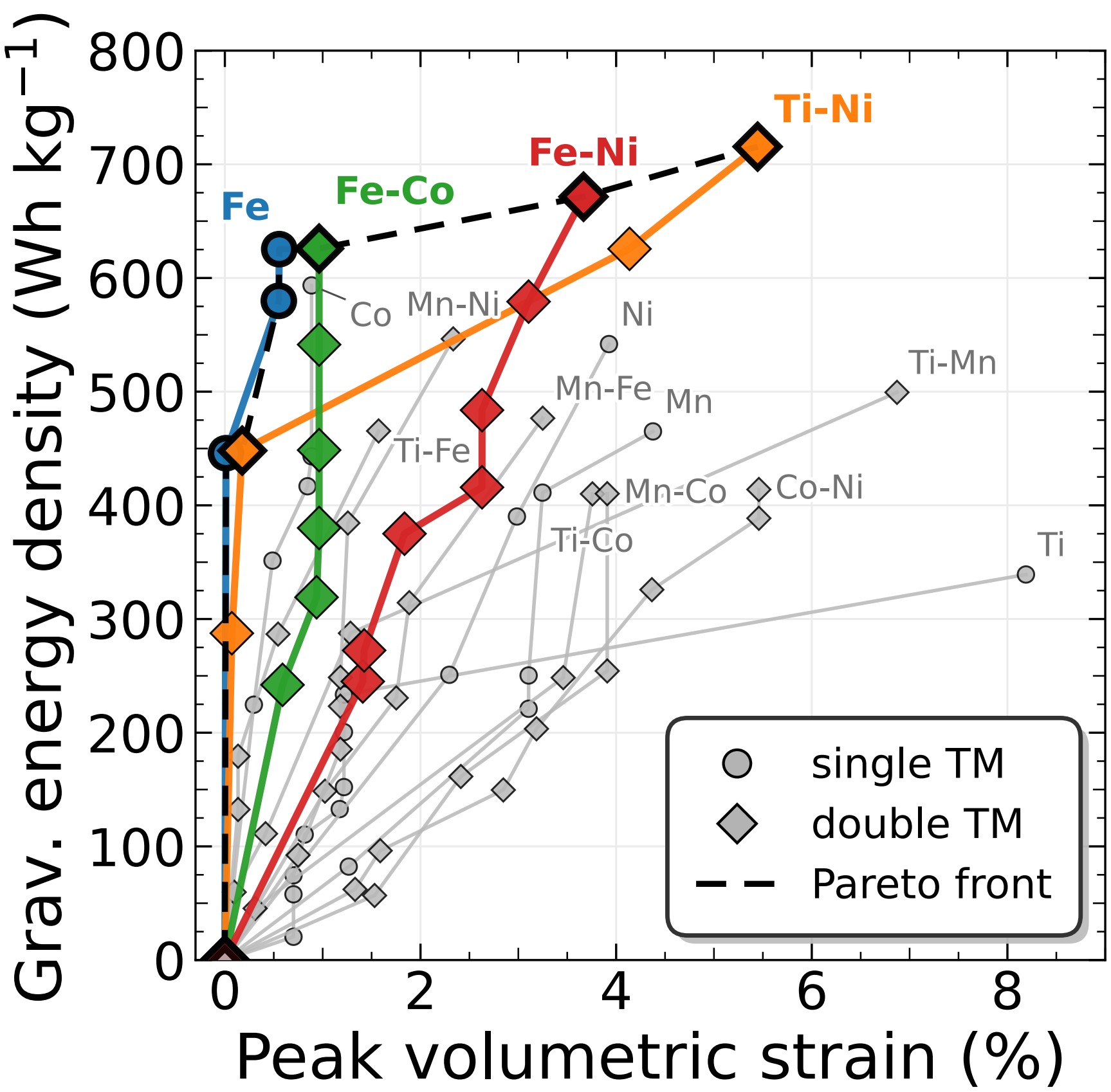

**Figure 6**. Estimated gravimetric energy density versus peak volumetric strain for all fifteen systems in **Figure 2** (circles: single TM; diamonds: double, mixed TM). We integrate the energy density over the computed voltage profile from x = 1 to the lowest-Na stable composition with x > 0, and we take the peak strain along the same desodiation path. The fully desodiated end member (x = 0) is excluded because of its large volume contraction. Colored markers with black outlines, connected by the dashed line, form the Pareto front.

**Figure 6** places four systems on the Pareto front (black-outlined markers and dashed line): $Na_xFeO_2$, $Na_xFe_{0.5}Co_{0.5}O_2$, $Na_xFe_{0.5}Ni_{0.5}O_2$, and $Na_xTi_{0.5}Ni_{0.5}O_2$. The front contains the non-dominated compositions, *i.e.*, those for which no other composition among the fifteen systems offers a higher energy density at equal or lower volumetric strain; every other composition is dominated by at least one point on the front. Thus, these four systems offer the best available compromise between energy density and volumetric strain. Note that the Pareto front ranks compositions by energy density and strain alone and does not account for whether they can be reached reversibly in real experiments.

We use **Figure 6** to identify which transition-metal combinations appear feasible in terms of volume strain and energy density. In general, Fe and Ni carry the highest-voltage redox couples in both the single-TM and mixed-TM oxides (**Figure 4a**), so Fe- and Ni-based systems achieve the largest integrated energy densities. $Na_xFeO_2$ retains O3 stacking with only 0.6% strain, whereas the P3-forming $Na_xFe_{0.5}Co_{0.5}O_2$, $Na_xFe_{0.5}Ni_{0.5}O_2$, and $Na_xTi_{0.5}Ni_{0.5}O_2$ accumulate 1.0%, 3.7%, and 5.5% strain, respectively. Strains of this magnitude matter in practice because they increase voltage hysteresis, limit rate capability, and drive capacity fade[30].

However, Fe-based layered materials come with a caveat. $Fe^{4+}$ formed at deep states of charge is prone to irreversible migration into the Na layer and therefore limits the material's theoretical energy density[13,42,52]. As described in the Results, the reversible cycling of $Na_xFeO_2$ is limited to x > 0.5, and the reversible window extends only to x ≈ 0.3 even when Fe is mixed with Co or Ni (x ≈ 0.30 for $Na_xFe_{0.5}Co_{0.5}O_2$ and x ≈ 0.34 for $Na_xFe_{0.5}Ni_{0.5}O_2$), since Co and Ni are oxidized before Fe, and thus only delay the formation of $Fe^{4+}$ [42,45]. In contrast, the curves of **Figure 6** continue to x = 1/3 for $Na_xFeO_2$ and to x = 1/4 for the two TM-mixed systems. When the curves are ended at the stable compositions closest to practical limits (x = 1/2 for $Na_xFeO_2$ and x = 1/3 for the two mixtures), the energy density decreases from 625 to 446 Wh $kg^{-1}$ for $Na_xFeO_2$, from 626 to 541 Wh $kg^{-1}$ for $Na_xFe_{0.5}Co_{0.5}O_2$, and from 672 to 579 Wh $kg^{-1}$ for $Na_xFe_{0.5}Ni_{0.5}O_2$. In **Figure 6**, $Na_xTi_{0.5}Ni_{0.5}O_2$ appears particularly promising: the two-electron $Ni^{2+}/Ni^{4+}$ couple provides both a high voltage and, with ½ Ni per formula unit, the full Na capacity, while $Ti^{4+}$ remains redox-inactive and Jahn–Teller-silent[47]. However, the accumulated strain in $Na_xTi_{0.5}Ni_{0.5}O_2$, which remains below 0.2% down to x = ½, increases abruptly to 5.5% in the second half of the charge, so extracting the full Na content is likely difficult in practice. Consistently, only about half of the Na is extracted in experiment (**Figure 2**), which limits the energy density to 448 Wh kg$^{-1}$ at $x$ = ½.

Except for $Na_xMn_{0.5}Ni_{0.5}O_2$, in which Mn remains JT-silent $Mn^{4+}$, Mn-rich systems are displaced from the Pareto front by the Jahn–Teller strain shown in **Figure 5**. Suppressing the JT distortion in $Na_xMn_{0.5}Ni_{0.5}O_2$ minimizes the strain penalty in **Figure 6** while retaining a competitive gravimetric energy density, placing this material just behind the Pareto front. Furthermore, when the Fe-based systems are limited to their reversible windows, the energy density of $Na_xMn_{0.5}Ni_{0.5}O_2$ (546 Wh kg$^{-1}$ at x = 1/3) is comparable to those of $Na_xFe_{0.5}Co_{0.5}O_2$ (~541 Wh kg$^{-1}$) and $Na_xFe_{0.5}Ni_{0.5}O_2$ (~579 Wh kg$^{-1}$). Thus, $Na_xMn_{0.5}Ni_{0.5}O_2$ stands beside $Na_xTi_{0.5}Ni_{0.5}O_2$ as one of the best practical choices. Its $Mn^{4+}$ lock keeps the accumulated strain at 2.3%; it is among the most widely cycled O3 cathodes[41,53,54], and it is the basis of the Mn-Ni-based multi-TM and high-entropy compositions[32,64,65].

Finally, it would be useful to predict the stable phase at each Na content from a simple descriptor, without costly electronic-structure calculations. For as-synthesized materials, two descriptors have been proposed to distinguish O3 from P2 phases: (i) the slab-anisotropy ratio, i.e., the interslab distance relative to the slab thickness, *d*(O–Na–O)/*d*(O–TM–O),[19,31] and (ii) the cationic potential[18].

However, it is unclear whether these descriptors remain valid during cycling, where the O3 and P3 phases, connected by slab gliding (**Figure 1**), compete at every Na content, with energy differences of 2–171 meV/f.u. at the stable compositions. Using our dataset of single- and mixed-TM layered oxides, we test both descriptors against the O3/P3 competition at each state of charge (**Figure 7**). We begin with the slab-anisotropy ratio (**Figure 7a**). An expanded Na layer is a geometric prerequisite for P3: the prismatic coordination places the oxygen layers of adjacent slabs directly above each other across the Na layer, so these oxygen layers are no longer close-packed, and a trigonal prism of spheres is taller than an octahedron of the same spheres[24].

Consistently, all P3 ground states in **Figure 7a** cluster at large *d*(O–Na–O) and high anisotropy ratios. However, many O3 states occupy the same region, so an expanded interlayer is necessary but not sufficient for P3 stability. Whether P3 actually prevails over O3 depends on the Na–vacancy ordering energy and is realized kinetically through slab glide[24,25,30].

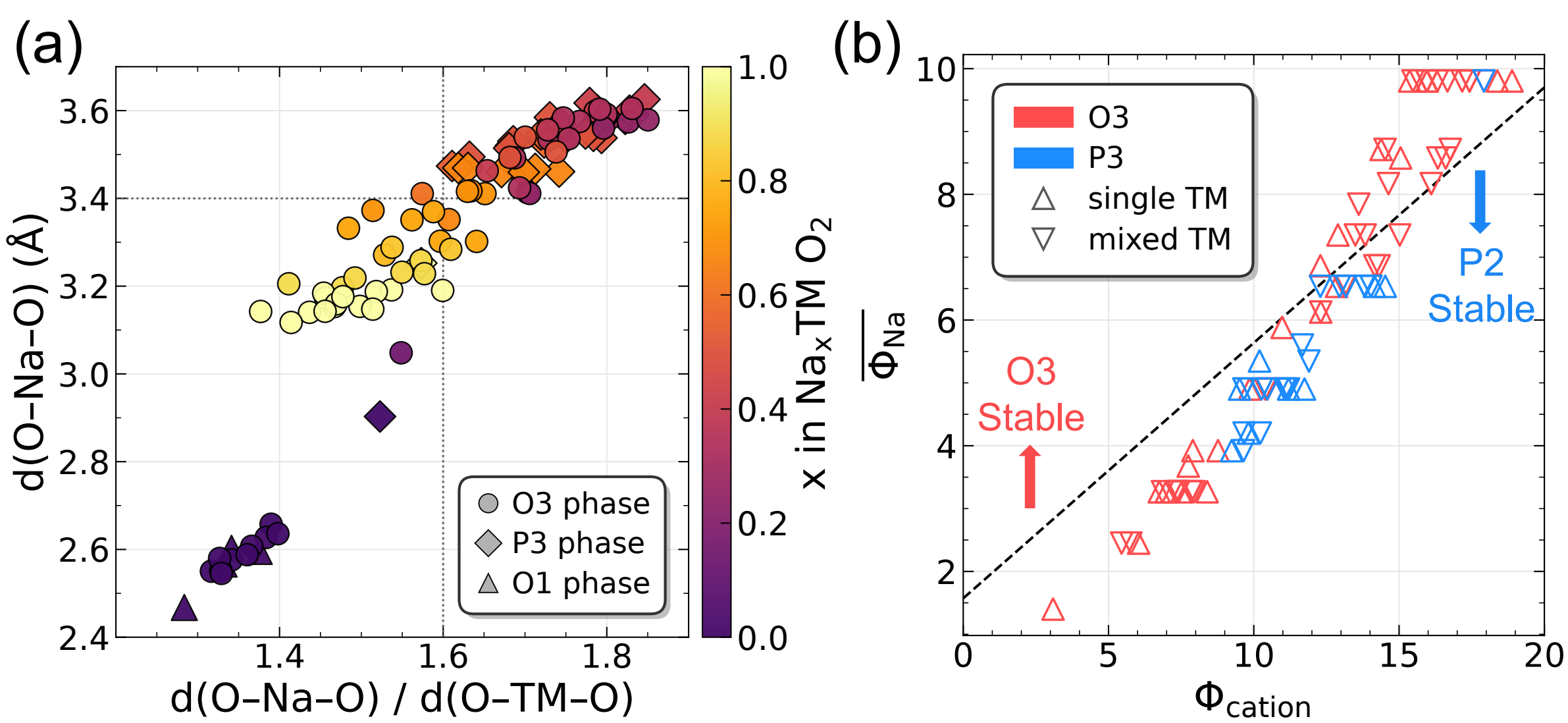

**Figure 7**. Descriptors of O3-versus-P3 stacking selection for thermodynamically stable compositions. (a) Slab-anisotropy ratio d(O-Na-O)/d(O-TM-O) plotted against the Na-slab thickness d(O-Na-O); marker shape = stable stacking (O3 circle, P3 square, O1 triangle), color = Na content x. (b) Cationic-potential map after[18]: Phi_cation = (Phi_TM x Phi_Na)/Phi_O versus Phi_Na = x/R(Na); marker color = stable stacking (O3 red, P3 blue), shape = single (up triangle) versus mixed (down triangle) TM; dashed line = empirical O3/P3 boundary.

Next, we test the cationic potential, which condenses each composition into a single electrostatic scalar, $\Phi_{cation} = \Phi_{TM}\Phi_{Na}/\Phi_{O}$, where $\Phi = n/R$ is the ionic potential of an ion (its charge *n* divided by its ionic radius *R*), $\Phi_{TM}$ is averaged over the transition metals, and $\Phi_{Na}$ is weighted by the Na content. This descriptor was originally proposed to predict P2-versus-O3 stacking at synthesis[18]. We define these quantities on the basis of the oxidation states from **Figure 3** and the "empirical" Shannon radii[66]. It recovers the qualitative trend that a high $\Phi_{cation}$ favors O3. However, although its empirical boundary classifies as-synthesized O3/P2 compounds well, it assigns only ~64% of the O3/P3 ground states here correctly, slightly below the 67% obtained by always predicting the majority class (O3). Furthermore, its two axes, $\Phi_{cation}$ and $\Phi_{Na}$, are nearly collinear ($r \approx 0.97$) because both depend on the Na content, so the map is effectively one-dimensional.

In our view, the main limitation of scalar descriptors, such as $\Phi_{cation}$, is that they encode neither the Na–vacancy ordering nor its electrostatic energy. A scalar descriptor takes a single value per composition, an apparently drastic reduction of dimensionality here. In contrast, the stable phase is determined by the energies of the most stable Na–vacancy orderings of the O3 and P3 phases at that composition. The O3/P3 boundary is thus set by ordering energies of a few meV/atom, which no average of radii and valences can resolve; nor does the cationic potential distinguish prismatic from octahedral Na sites. State-of-charge-resolved predictions may require explicit enumeration of Na–vacancy orderings (and, in mixed-TM compounds, TM orderings), as performed in this work, together with glide energetics[67,68]. A charge-resolved descriptor, such as an anionic potential derived from computed (*e.g.*, Bader, Mulliken, *etc.*) charges of the actual ground-state structures, offers another possible route that is not tested here.

## Conclusion

In this work, we presented, to our knowledge, the first consistent, ordering-resolved map of the O3/P3/O1 family of sodium layered oxides. A single library of 41 Na/vacancy orderings, drawn from existing structures, was applied uniformly to five single-transition-metal (TM) $Na_xTMO_2$ and all ten mixed $Na_xTM_{0.5}TM'_{0.5}O_2$ oxides (TM = Ti, Mn, Fe, Co, Ni), with Na-layer stacking, TM ordering, and TM-layer stacking treated explicitly. Screening more than 120,000 candidate structures with a machine-learning interatomic potential and performing 1,025 DFT+*U* calculations, we derived, for each system, the phase diagram, voltage profile, lattice evolution, TM oxidation states, and octahedral distortions. The resulting map provides a single, comparable reference spanning many commercially relevant compositions, from which three general rules emerge:

(i) **The redox ladder is largely independent of the partner metal.** Over the first half of charge, the redox couples order $Ti^{3+/4+}$ < $Mn^{3+/4+}$ < $Co^{3+/4+}$ < $Ni^{2+/3+}$ < $Fe^{3+/4+}$. This ordering explains the charge transfer in the pristine compounds, the order in which the metals oxidize, and why Fe is always oxidized last. It also correctly predicts the redox sequence of compositions not computed here, such as $NaNi_{1/3}Fe_{1/3}Mn_{1/3}O_2$.

(ii) **Jahn–Teller distortions tend to be cooperative, but are often ameliorated in mixed transition metal compositions.** Strong distortions are confined to $Mn^{3+}$, $Ni^{3+,}$ and low-spin $Co^{2+}$. About 40% of the distortion in the end members arises from neighboring octahedra elongating along the same axis. Mixing transition metals suppresses it in two ways: charge transfer mitigates the effects of Jahn–Teller-active ions, and Jahn–Teller-inactive partners dilute them and break the cooperative alignment.

(iii) **O3/P3 competition depends on the Na/vacancy ordering.** Across the stable compositions, O3 and P3 differ by 2–171 meV/f.u., so which stacking wins depends on the Na/vacancy ordering energy. The cationic potential classifies only ~64% of the O3/P3 ground states correctly, below the 67% obtained by always predicting O3. The interslab distance is necessary for P3 but not sufficient. Predicting the stacking at each state of charge therefore requires explicit enumeration.

From this analysis, we constructed an energy-density versus strain Pareto front comprising $Na_xFeO_2$, $Na_xFe_{0.5}Co_{0.5}O_2$, $Na_xFe_{0.5}Ni_{0.5}O_2$, and $Na_xTi_{0.5}Ni_{0.5}O_2$. Because $Fe^{4+}$ migration limits the Fe-based systems in practice, $Na_xTi_{0.5}Ni_{0.5}O_2$ and $Na_xMn_{0.5}Ni_{0.5}O_2$ emerge as the most practical choices. We released all ordering-resolved structures and energies openly as a reference for quantifying entropy, modeling kinetics, and training machine-learning interatomic potentials. The approach extends naturally to layered oxides with more than two transition metals, including high-entropy compositions.

## Methodology

**Ordering settings.** The three stackings (O1, O3, and P3) investigated here differ in how the Na sites connect to the $TMO_2$ slabs (**Figure 1**). In O1, each octahedral Na site shares faces with $TMO_6$ octahedra in both the slab above and the slab below; in O3, a Na site shares only edges with the neighboring slabs; and in P3, each prismatic Na site shares a face with one $TMO_6$ octahedron in one slab and edges with three octahedra in the other. Therefore, in the O1 and the O3 phases, the Na sites form a triangular lattice with one site per $TMO_2$ formula unit, while in P3 the two prism orientations form a honeycomb lattice with two sites per formula unit, of which only one can be occupied since adjacent prisms share a face and are separated by only ~1.7 Å [23]. Each Na/vacancy ordering in this work is an in-plane decoration of the triangular or honeycomb Na lattice at a fixed Na content.

The ordering library is compiled from previous computational and experimental studies, drawn in **Figures S1** and **S2** with labels that encode their provenance. Seventeen O3 orderings are the DFT ground states of the single-TM $Na_xTMO_2$ family identified by Toumar *et al.*[35], where AJT_j at x = 5/8 is excluded because its Na ions sit away from the ideal octahedral sites and no exact structural data are available. Four O3 and six P3 orderings originate from Kaufman and Van der Ven in $Na_xCoO_2$ and $Na_xCrO_2$[23,36]. The experimentally refined orderings are two O'3 orderings of $Na_xMnO_2$ at x = 1/3 and 1/2 [27], the $Na_{5/8}MnO_2$ superstructure[26], the $Na_{1/2}VO_2$ superstructure[34], and five $Na_xNiO_2$ orderings including the monoclinic O'3 $NaNiO_2$ end member[29,37]. Orderings that duplicate another one up to a monoclinic distortion are cross-referenced in **Figures S1** and **S2**. Together with the fully sodiated and fully desodiated O3 and P3 hosts and the desodiated O1 host, the library includes 41 structures, which are applied uniformly to every TM chemistry.

Since an in-plane ordering can be shifted laterally between successive Na layers, which changes the interlayer interactions, the relative layer stacking is treated as an explicit degree of freedom following Toumar et al.[35]. Each ordering is built in a supercell with three Na layers; the Na/Vac ordering of a reference layer is propagated to the other two layers by every possible in-plane translation, and symmetry-equivalent stackings are removed. Candidate stacking models were relaxed with the MACE-OMAT-0 (medium) machine-learning interatomic potential[69,70] until residual forces fell below 0.05 eV/Å and then ranked by energy; only the lowest-energy stacking of each Na–vacancy ordering was passed to the DFT workflow.

Mixed-TM oxides have two additional degrees of freedom: in-plane TM ordering and TM-layer stacking. Because the two metals are mixed 1:1, cells with an odd number of TM sites per layer were doubled in-plane. We ranked in-plane orderings by Ewald energy and retained the five lowest; for each, we selected TM-layer stacking by single-point MACE-OMAT-0 energies before selecting Na-layer stacking as described above. The lowest DFT energy at each composition was used to construct the convex hull.

**Density Functional Theory settings.** All structures were relaxed with VASP[71,72] using the generalized gradient approximation, PBE[73], the Hubbard *U* corrections[74–76] and the PAW potentials (PBE_54 library)[77,78] as suggested by the Materials Project MPRelaxSet[79,80]. All structures were relaxed with a plane-wave cutoff of 680 eV, a *k*-point spacing of 0.3 $Å^{-1}$, the DFT-D3(BJ) dispersion correction, and spin polarization initialized ferromagnetically, until residual forces fell below 0.02 eV/Å and the electronic convergence criterion was $10^{-6}$ eV per atom. We referenced formation energies per $TMO_2$ formula unit to the fully sodiated and fully desodiated end-member compositions of each system, and we obtained the convex hull and voltage profile from hull-stable compositions, using the DFT energy of the body-centered cubic Na phase (–1.31 eV per atom) as the reference. Oxidation states were assigned to each TM ion by matching its on-site magnetic moment to the expected range for each oxidation and spin state, with ambiguous cases resolved by charge neutrality, and then averaged per element (**Figure 3**). We evaluated energies above the Materials Project convex hull[79] by re-relaxing the hull-stable structures with the default MPRelaxSet settings (520 eV cutoff), applying the Materials Project 2020 compatibility corrections[83], and building the phase diagram alongside the Materials Project entries for each chemical system.

**Van Vleck analysis.** The Van Vleck modes of each $TMO_6$ octahedron in the hull-stable structures were calculated with the VanVleckCalculator package[62] as $Q_2 = (l_y - l_x)/2$ and $Q_3 = (2l_z - l_x - l_y)/(2\sqrt{3})$, where $l_x$, $l_y$, and $l_z$ are the O–O distances of the octahedron along the x, y, and z axes. For each structure, one set of orthogonal axes is used for all the octahedra, which is obtained by averaging the O–TM–O directions over the octahedra, and z is assigned to the elongated axis of the most elongated octahedron, so that the angle $\varphi$ can be compared within a structure and among the structures. The oxidation states in **Figure 5** are the atom-by-atom assignments from the magnetic moments described above, applied to the single-TM and mixed-TM oxides.

## Acknowledgments

We are grateful to the Research Computing Data Core at the University of Houston for assistance with the calculations presented in this work. P. C. acknowledges the Welch Foundation under grants L-E-001-19921203, and E-2227-20250403. H. H. L. and J. D. B. acknowledge support from the Welch Foundation under grant E-2179-20240404.

## Conflicts of interest

All authors declare no competing interests.

## Data availability

Computational data related to this study, including Python scripts, input, and output files of the simulations, are available on GitHub at https://github.com/caneparesearch/project_NLO_DFT

**—Supplementary Information—**

# Decoding Composition–Structure–Performance Relations in Sodium Layered Oxide Positive Electrode Materials

Yunyeong Choi[1,2], Hasindhi H. Lokuge[1,3], Joshua D. Bocarsly[1,3], and Pieremanuele Canepa[1,2,4*]

[1]Texas Center for Superconductivity at the University of Houston, Houston, TX 77204, USA
[2]Department of Electrical and Computer Engineering, University of Houston, Houston, TX 77204, USA
[3]Department of Chemistry, University of Houston, Houston, TX 77204, USA
[4]Department of Mechanical Engineering, Purdue University, West Lafayette, IN 47907, USA

*e-mail: canepa@purdue.edu

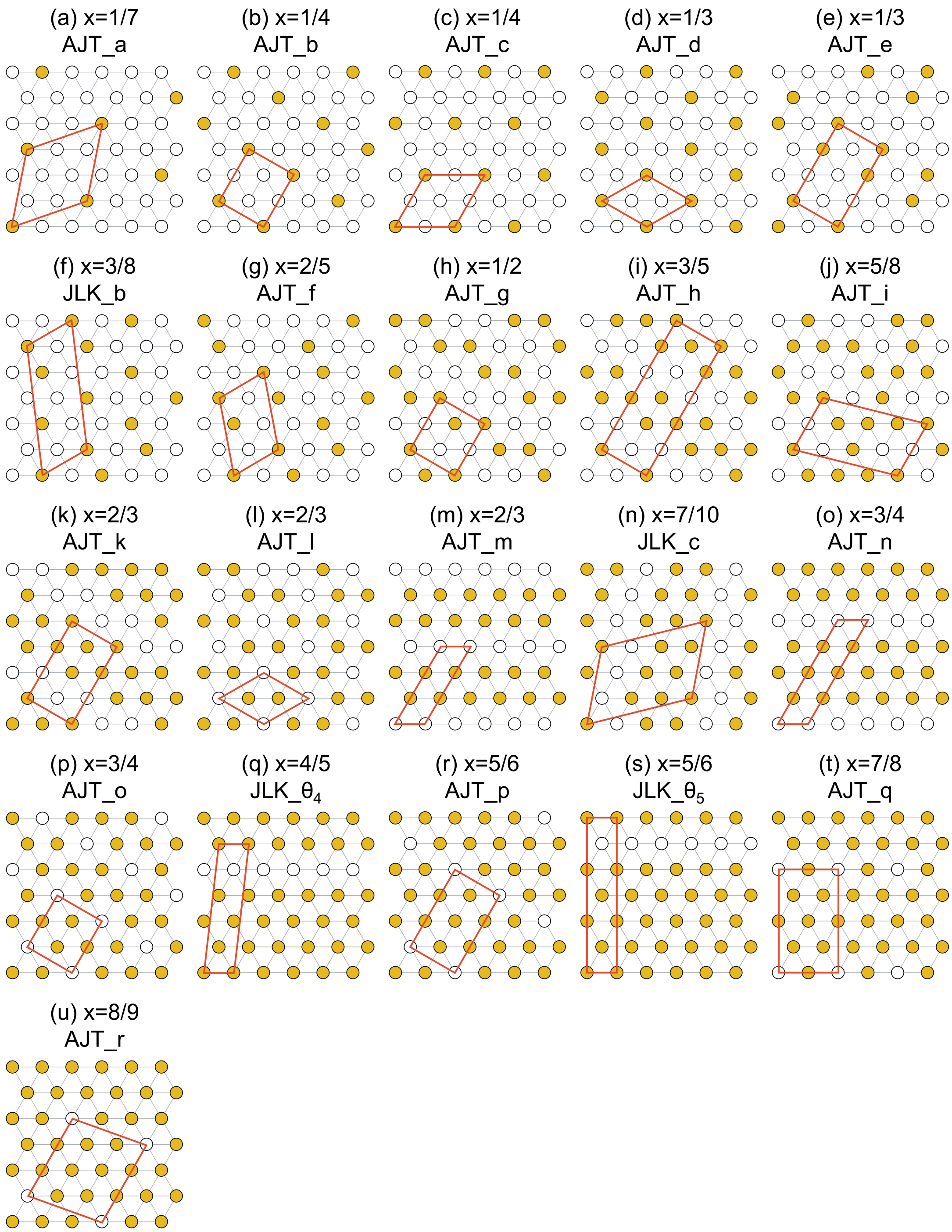

**Figure S1.** Na–vacancy ordering diagrams of the O3-host orderings used in the DFT calculations. Labels of each ordering encode provenance: the initials of the authors of the source study, followed by the ordering label used in that work, or by an alphabetic index where the source did not label its orderings (AJT: Toumar et al. 2015[1]; JLK: Kaufman & Van der Ven 2019 and 2022[2,3]; JMAS: Steele et al. 2025 and 2026[4,5]; XL: Li et al. 2014[6]; XC: Chen et al. 2018[7]; CD: Didier et al. 2012.[8] Orderings identical to a listed entry up to a monoclinic lattice distortion are omitted: XC_a and JMAS_a are variants of AJT_d, JMAS_b of AJT_f, and CD_a of AJT_g. XC_b and XL_a are omitted because small off-site Na displacements take their Na–vacancy orderings off the ideal triangular lattice.

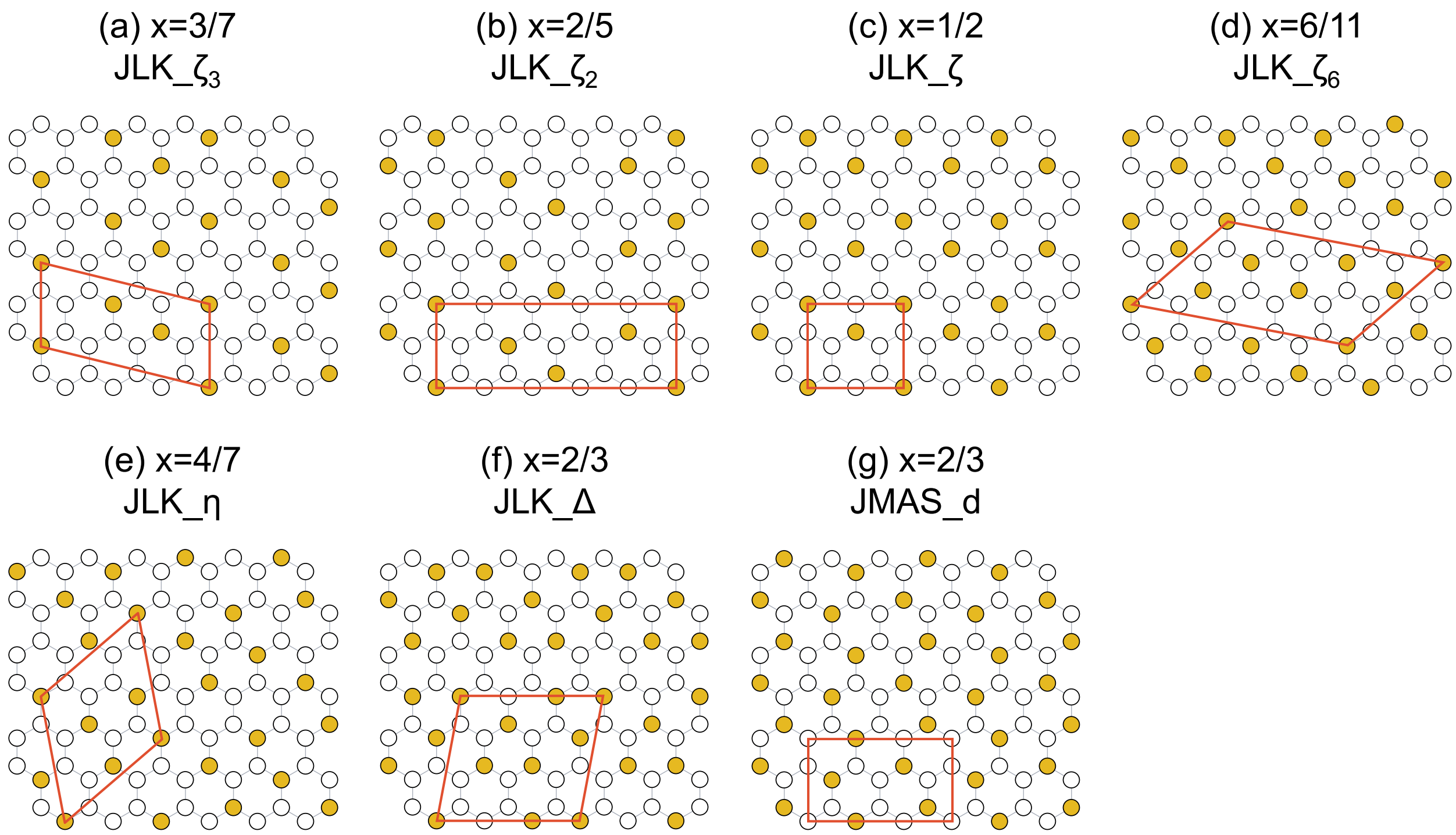

**Figure S2.** Na–vacancy ordering diagrams of the P3-host orderings used in the DFT enumeration. The naming convention follows Figure S1. Orderings identical to a listed entry up to a monoclinic distortion are omitted; JMAS_c is a variant of JLK_ζ.

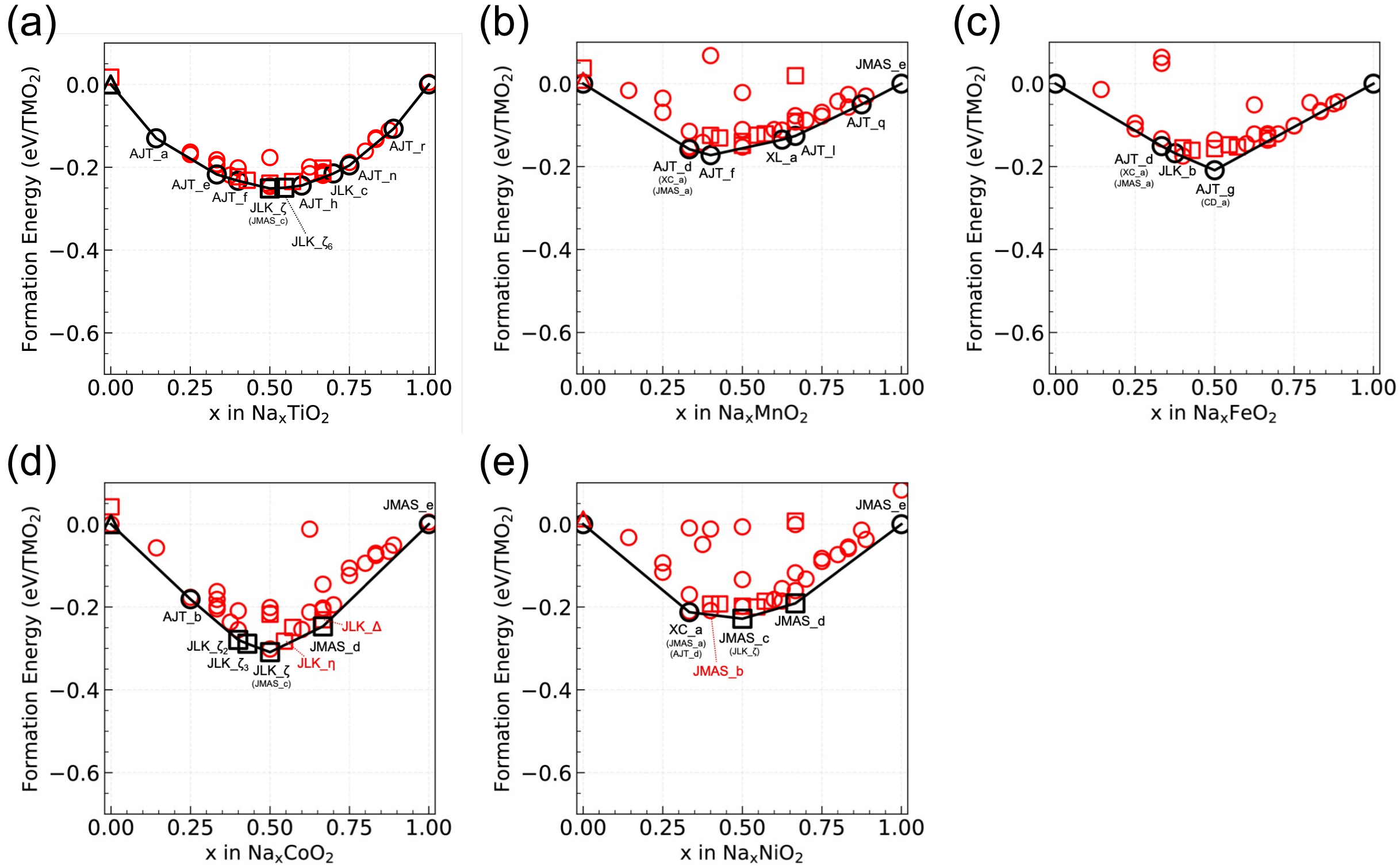

**Figure S3.** Convex-hull diagrams of the single-TM systems $Na_xTMO_2$ (TM = Ti, Mn, Fe, Co, Ni). Formation energy per $TMO_2$ formula unit, referenced to the tie-line between the most stable fully sodiated and fully desodiated endpoints (endpoints at ΔE = 0). Marker shape denotes the host stacking (circle O3, square P3, triangle O1); black-outlined markers are hull ground states, red-outlined markers lie above the hull.

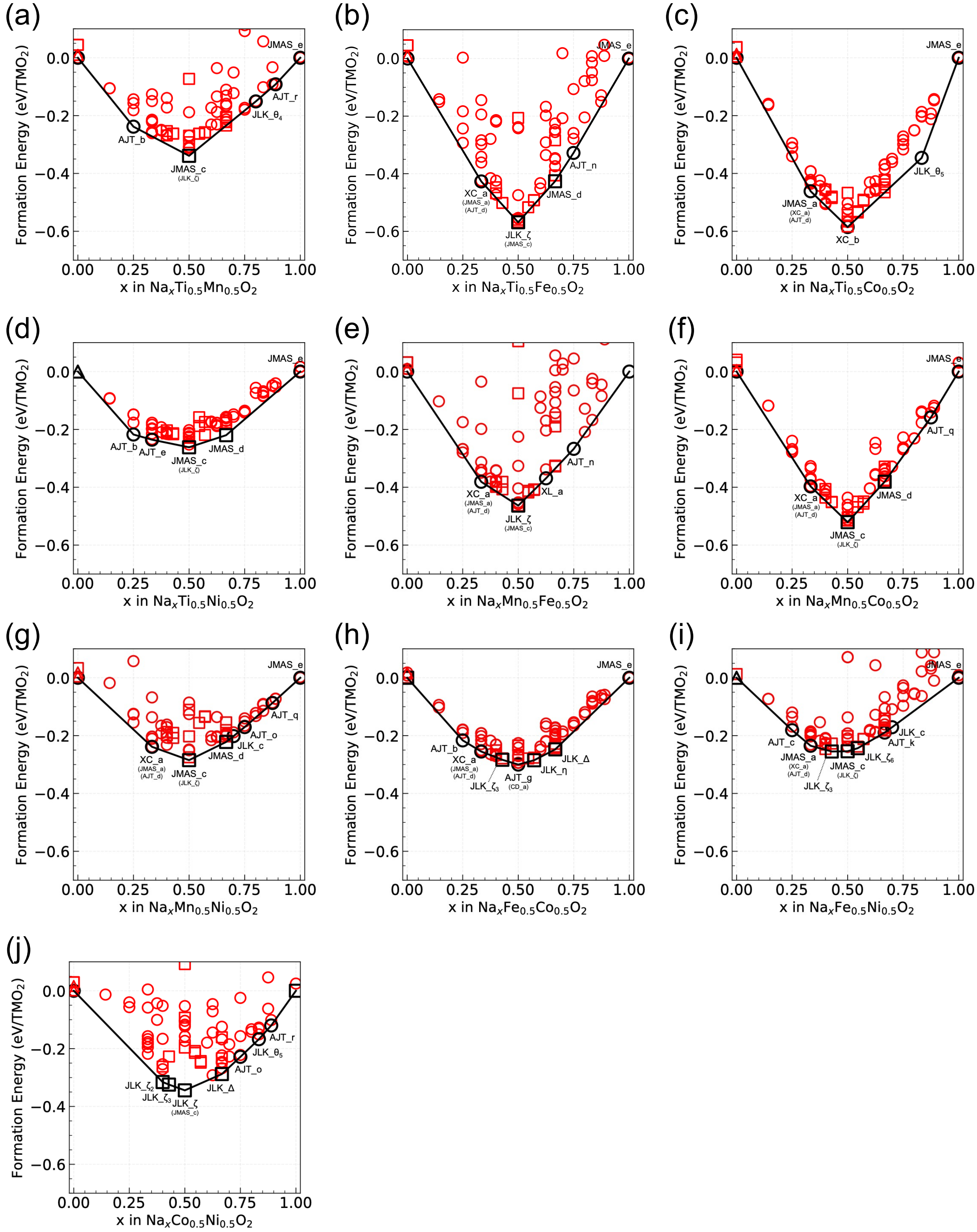

**Figure S4.** Convex-hull diagrams of the ten mixed systems $Na_xTM_{0.5}TM'_{0.5}O_2$, plotted as in Figure S3.

**Table S1.** Transition-metal oxidation states in the mixed-TM systems from experiment and from the DFT magnetic moments of this work. x is the Na content in $Na_xTM_{0.5}TM'_{0.5}O_2$, and the first line of each entry gives the pristine (x = 1) state. Experimental assignments are taken from the studies listed below the table. DFT assignments are the per-element averages of **Figure 3** of the main text at the hull-stable compositions.

| System | Experimental probe | Oxidation states, experiment | Oxidation states, DFT (this work) | Agreement |
|---|---|---|---|---|
| $\mathbf{Na_xMn_{0.5}Ni_{0.5}O_2}$ | Ni, Mn K-edge XANES/EXAFS, ex situ[9] | When fully sodiated: $Mn^{4+}/Ni^{2+}$<br><br>$Ni^{2+} \rightarrow Ni^{3+} \rightarrow Ni^{4+}$ on charge, $Mn^{4+}$ inactive | When fully sodiated: $Mn^{4+}/Ni^{2+}$<br><br>$Ni^{2+} \rightarrow Ni^{3+} \rightarrow Ni^{4+}$ on charge, $Mn^{4+}$ inactive | Yes |
| $\mathbf{Na_xTi_{0.5}Ni_{0.5}O_2}$ | Ti L-, Ni L- and O K-edge XAS[10] | When fully sodiated: $Ti^{4+}/Ni^{2+}$<br><br>$Ni^{2+} \rightarrow Ni^{3+}$ for $1 \geq x \geq 0.5$, $Ti^{4+}$ inactive | When fully sodiated: $Ti^{4+}/Ni^{2+}$<br><br>$Ni^{2+} \rightarrow Ni^{3+} \rightarrow Ni^{4+}$ on charge, $Ti^{4+}$ inactive | Yes |
| $\mathbf{Na_xFe_{0.5}Ni_{0.5}O_2}$ | $^{57}$Fe Mössbauer, Fe K- and Ni L-edge XAS[10,11]. TEM-EELS $L_3/L_2$ ratios, ex situ[14] | When fully sodiated: $Fe^{3+}/Ni^{3+}$<br><br>XAS/Mössbauer: $Ni^{3+} \rightarrow Ni^{4+}$, Fe stays 3+ (Fe K-edge shift from LMCT)<br>EELS: $Fe^{3+} \rightarrow Fe^{4+}$ for $1 > x \geq 0.5$, Ni only partly | When fully sodiated: $Fe^{3+}/Ni^{3+}$<br><br>$Ni^{3+} \rightarrow Ni^{4+}$ by x = 0.5, $Fe^{3+}$ kept ($Fe^{3.67+}$ only at x = 2/3)<br>$Fe^{3.33+}$ for $x \leq 0.43$ | Yes (XAS, Mössbauer)<br>No (EELS and DFT of ref. f) |
| $\mathbf{Na_xFe_{0.5}Co_{0.5}O_2}$ | Co, Fe K-edge XANES, ex situ[12] | When fully sodiated: $Fe^{3+}/Co^{3+}$<br><br>$Co^{3+} \rightarrow Co^{4+}$ first ($1 \geq x \geq 0.55$), then $Fe^{3+} \rightarrow Fe^{4+}$ ($x \leq 0.67$) | When fully sodiated: $Fe^{3+}/Co^{3+}$<br><br>$Co^{3+} \rightarrow Co^{4+}$ by x = 0.5, $Fe^{3+}$ kept<br>$Fe^{3+} \rightarrow Fe^{4+}$ for $x < 0.5$ | Yes |
| $\mathbf{Na_xMn_{0.5}Fe_{0.5}O_2}$ | $^{57}$Fe Mössbauer, Fe, Mn K-edge XANES[13] | When fully sodiated: $Mn^{3+}/Fe^{3+}$<br><br>$Mn^{3+} \rightarrow Mn^{4+}$ at low voltage, $Fe^{3+} \rightarrow Fe^{4+}$ only at high voltage | When fully sodiated: $Mn^{3+}/Fe^{3+}$<br><br>$Mn^{3+} \rightarrow Mn^{4+}$ by x = 0.5, $Fe^{3+}$ kept<br>$Fe^{3+} \rightarrow Fe^{4+}$ for $x < 0.5$ | Yes |
| $\mathbf{Na_xCo_{0.5}Ni_{0.5}O_2}$ | TEM-EELS Ni, Co $L_3/L_2$ ratios, ex situ[14] | When fully sodiated: $Co^{3+}/Ni^{3+}$ (assigned with DFT)<br><br>Ni $L_3/L_2$ rises for $1 > x \geq 0.5$, $Co^{3+}$ kept | When fully sodiated: $Co^{4+}/Ni^{2+}$<br><br>$Ni^{2+} \rightarrow Ni^{3.8+}$ by x = 0.4 while Co reduced to ~3.4+<br>$Co^{4+}/Ni^{4+}$ at x = 0 | Partly<br>Ni-first redox agrees, pristine valences differ |

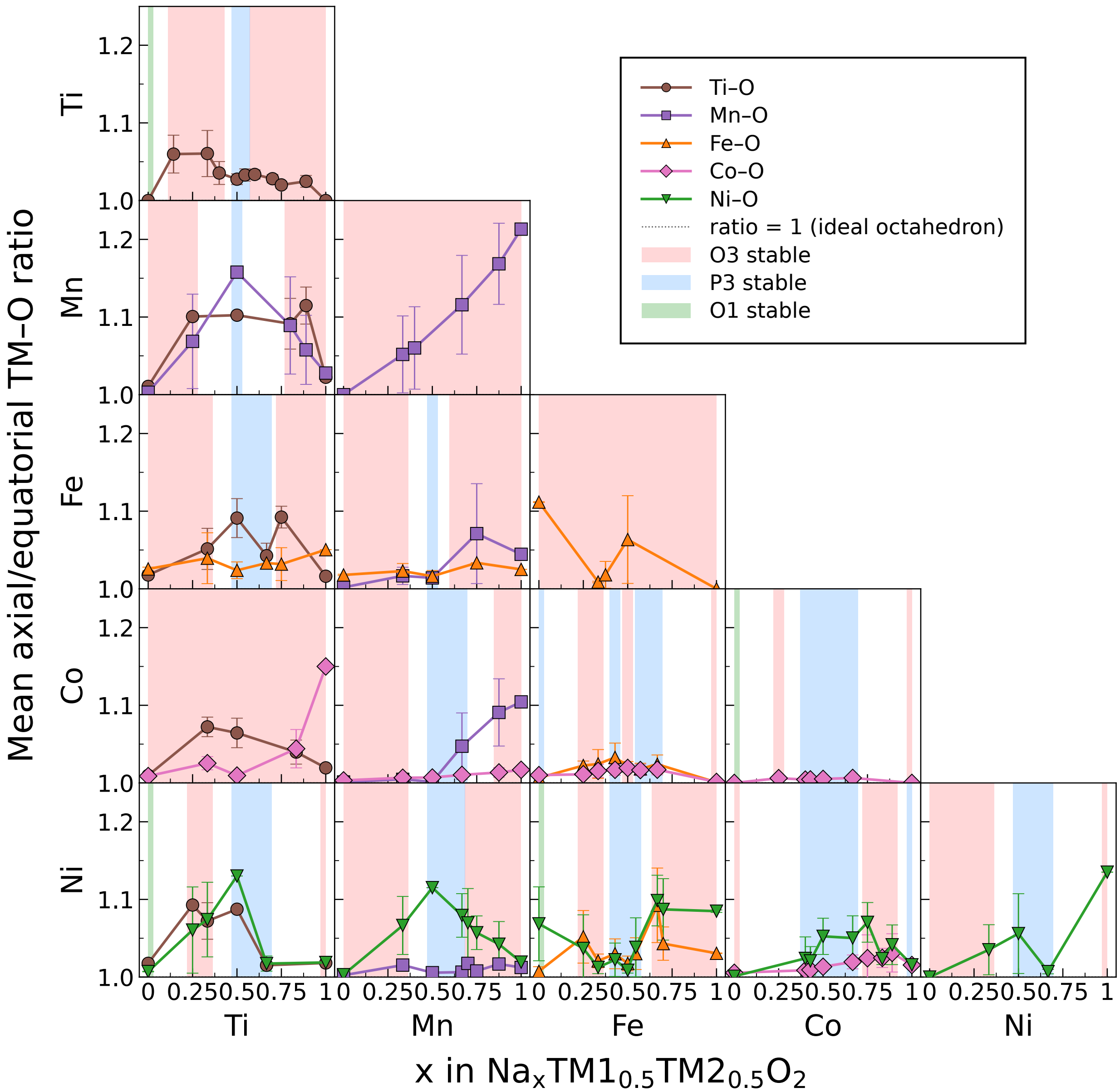

**Figure S5.** Jahn-Teller distortion of the $TMO_6$ octahedra versus Na content, quantified as the mean axial-to-equatorial TM-O bond-length ratio (two longest bonds = axial, four shortest = equatorial; 1.0 = ideal octahedron, >1 = JT elongation), averaged over all octahedra of each element. Error bars: standard deviation across the symmetry-distinct octahedra of that element. Markers color/shape-coded by element; shading: stable stacking (O3 red, P3 blue, O1 green).

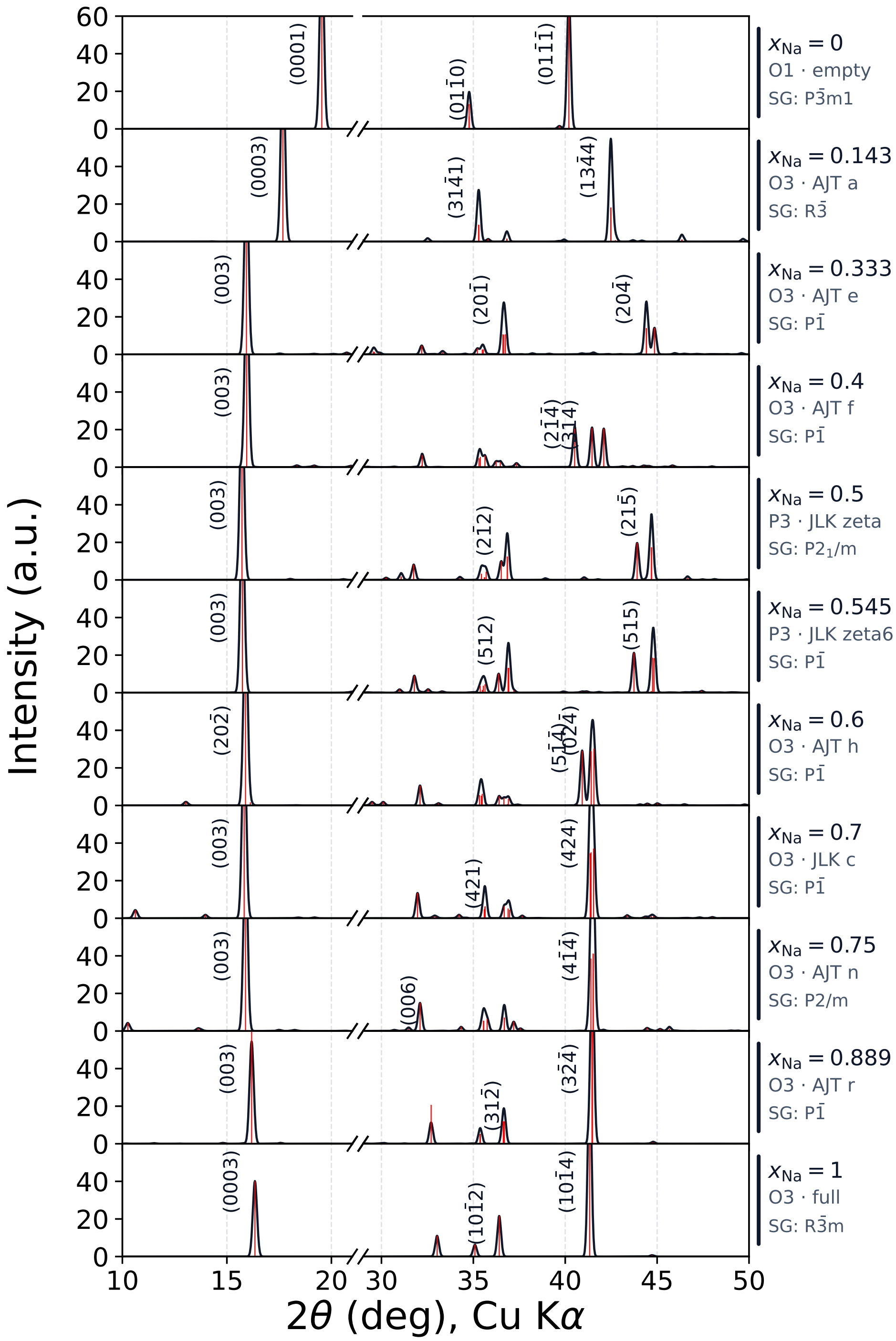

**Figure S6.** Simulated XRD patterns (Cu Kα) of every hull-stable structure of the single-TM systems of Figure S3, one system per page, rows ordered by Na content. Black curves: Gaussian-broadened profiles (σ = 0.10°, FWHM ≈ 0.24°); red sticks: underlying Bragg intensities; the three strongest reflections are labeled by their (hkl) indices; the 2θ axis is broken between 21° and 29°. Right-hand labels give the Na content $x_{Na}$, the stable stacking and Na–vacancy ordering labels, and the space group of the relaxed structure. (**a**) $Na_xTiO_2$.

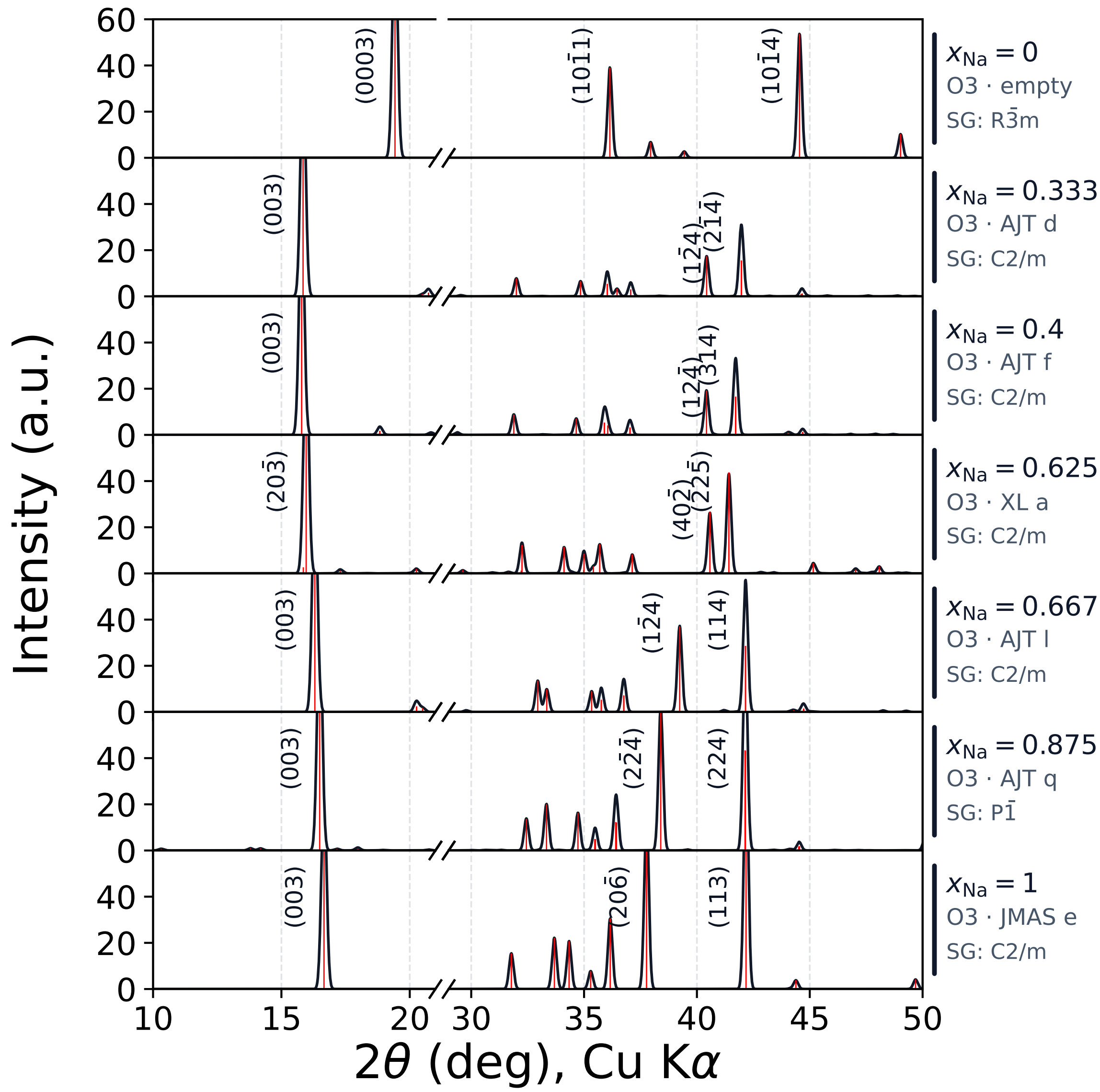

**Figure S6 (continued).** (**b**) $Na_xMnO_2$.

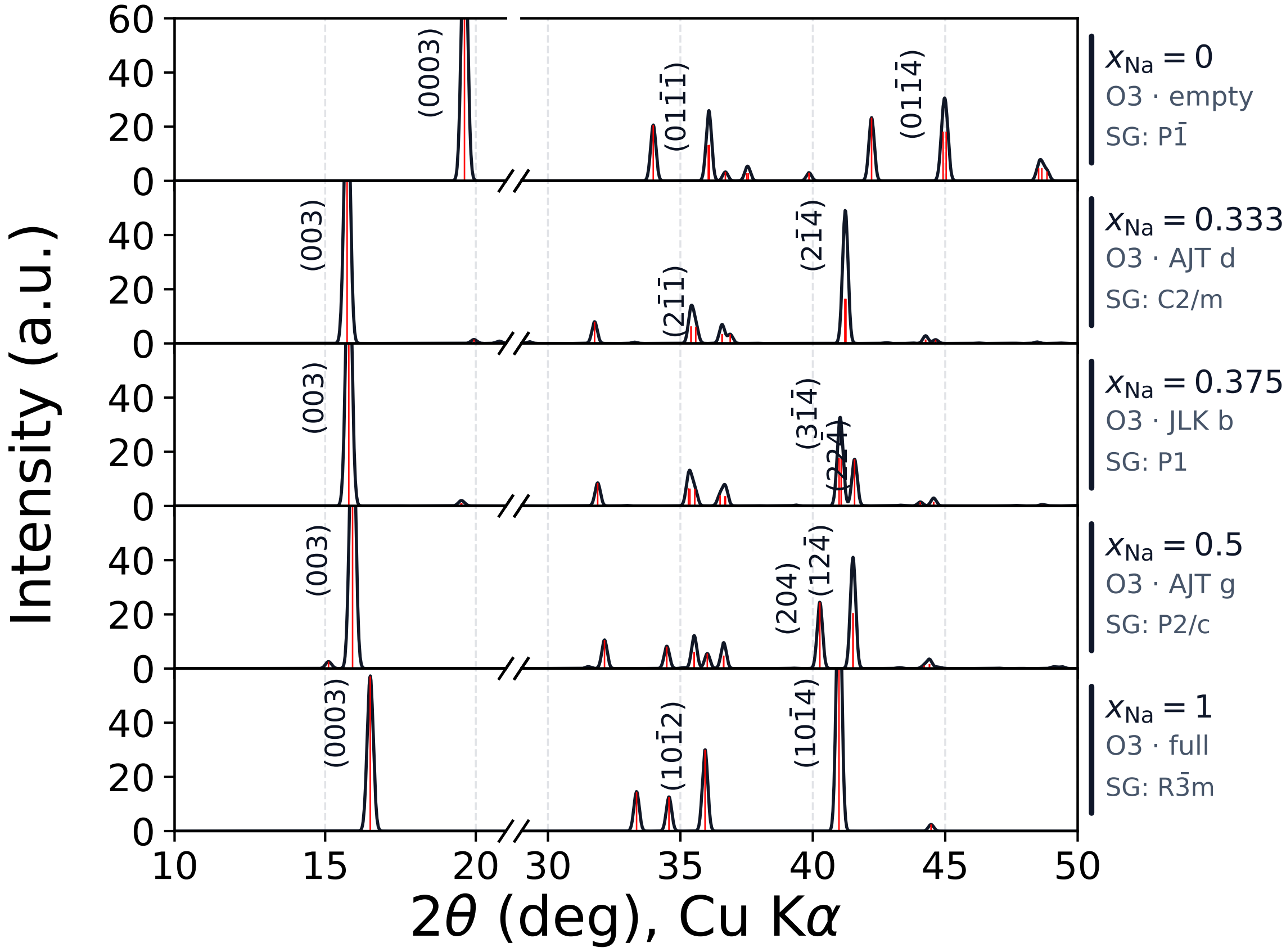

**Figure S6 (continued).** (**c**) $Na_xFeO_2$.

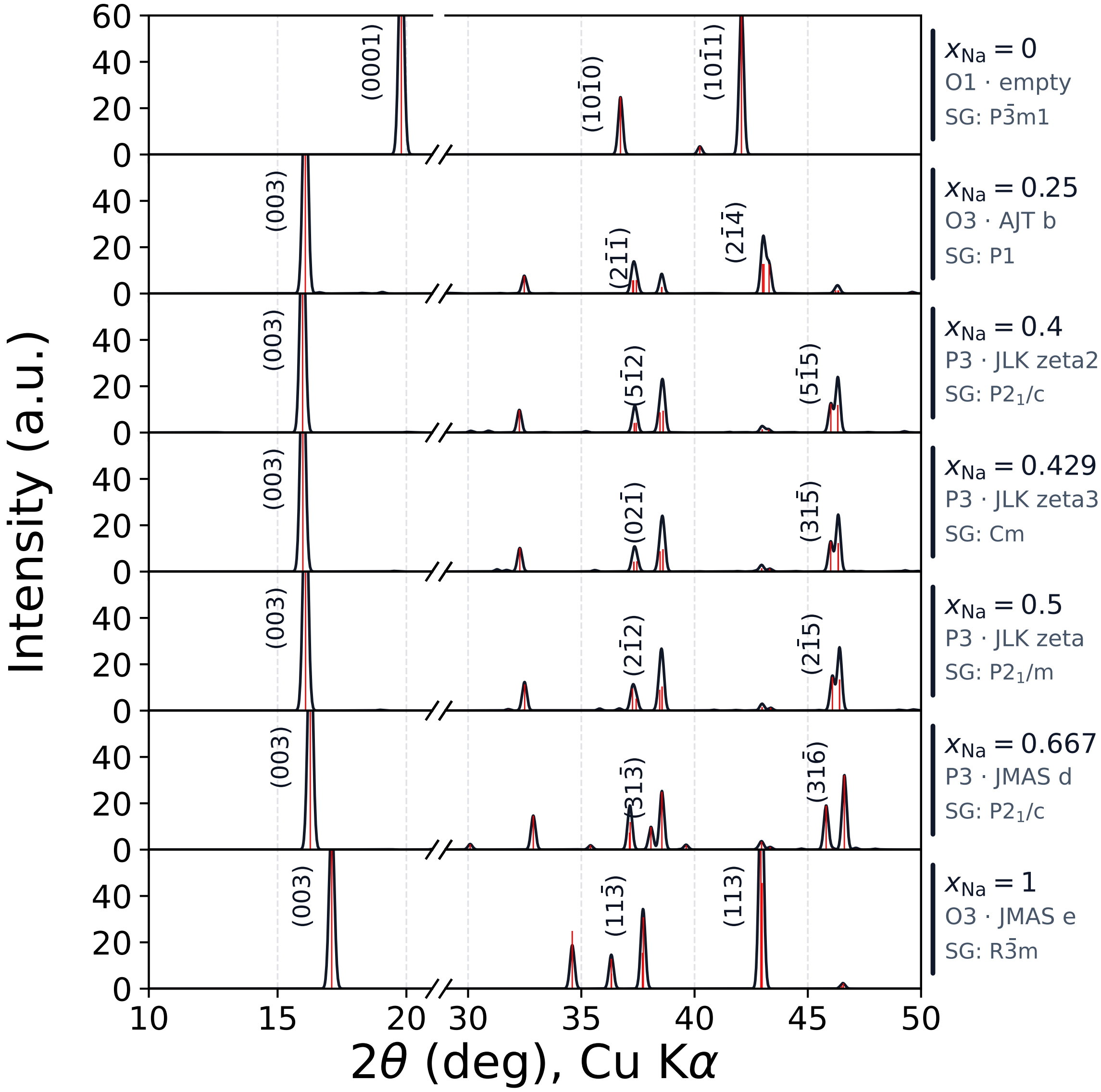

**Figure S6 (continued).** (**d**) $Na_xCoO_2$.

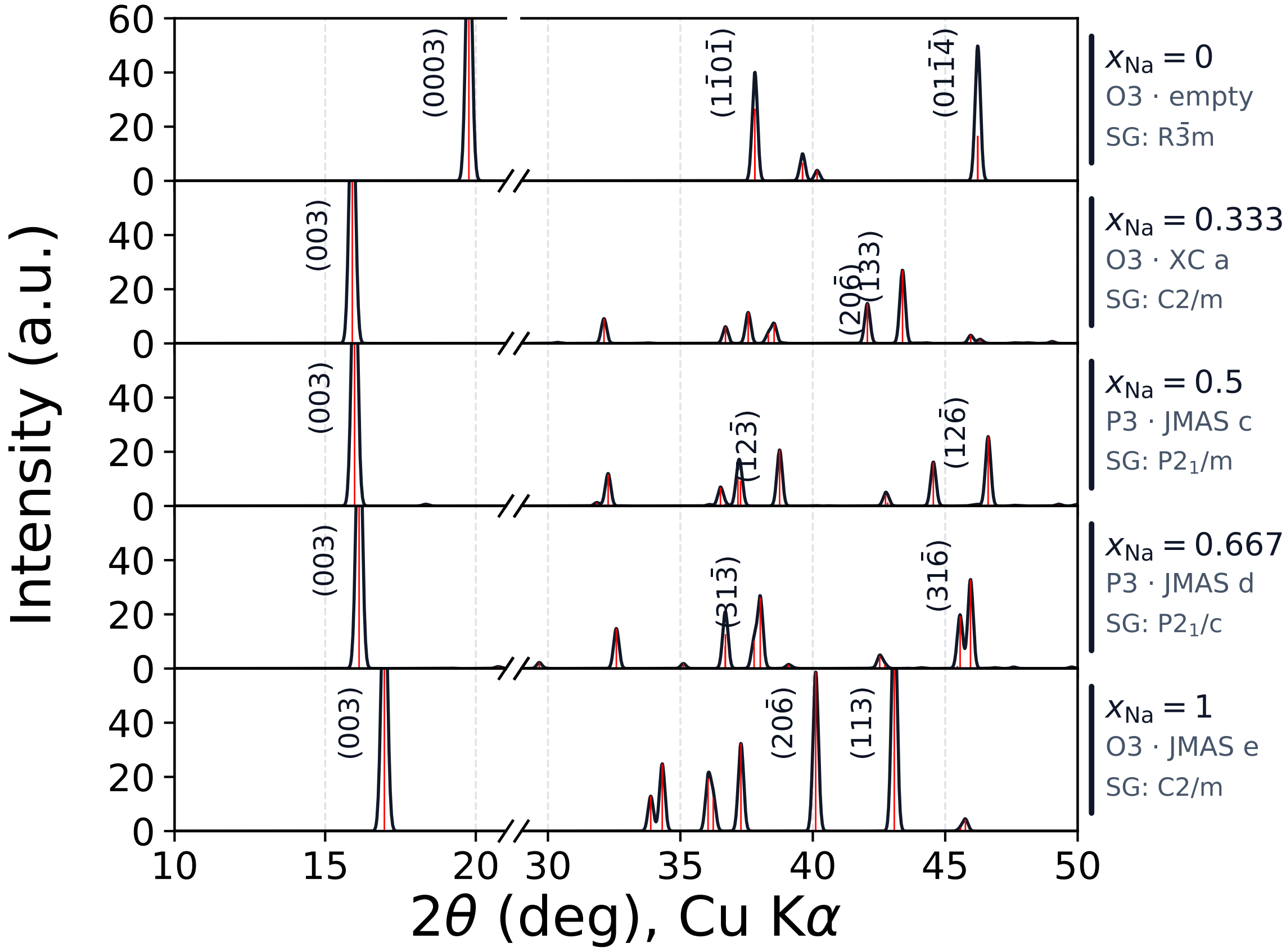

**Figure S6 (continued).** (**e**) $Na_xNiO_2$.

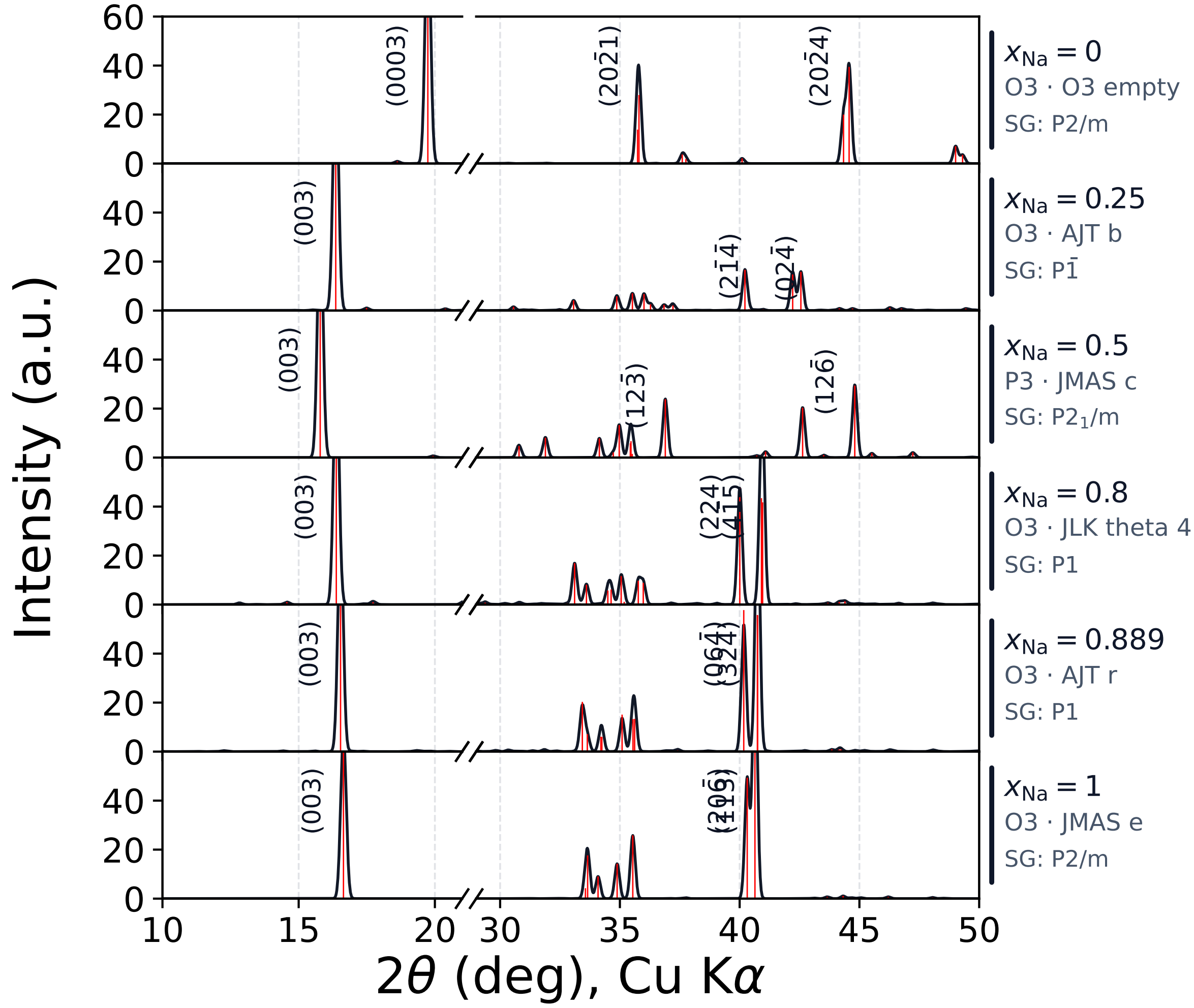

**Figure S7.** Simulated XRD patterns (Cu Kα) of every hull-stable structure of the ten mixed systems of Figure S4, presented as in Figure S6 (one TM pair per page). (**a**) $Na_xTi_{0.5}Mn_{0.5}O_2$.

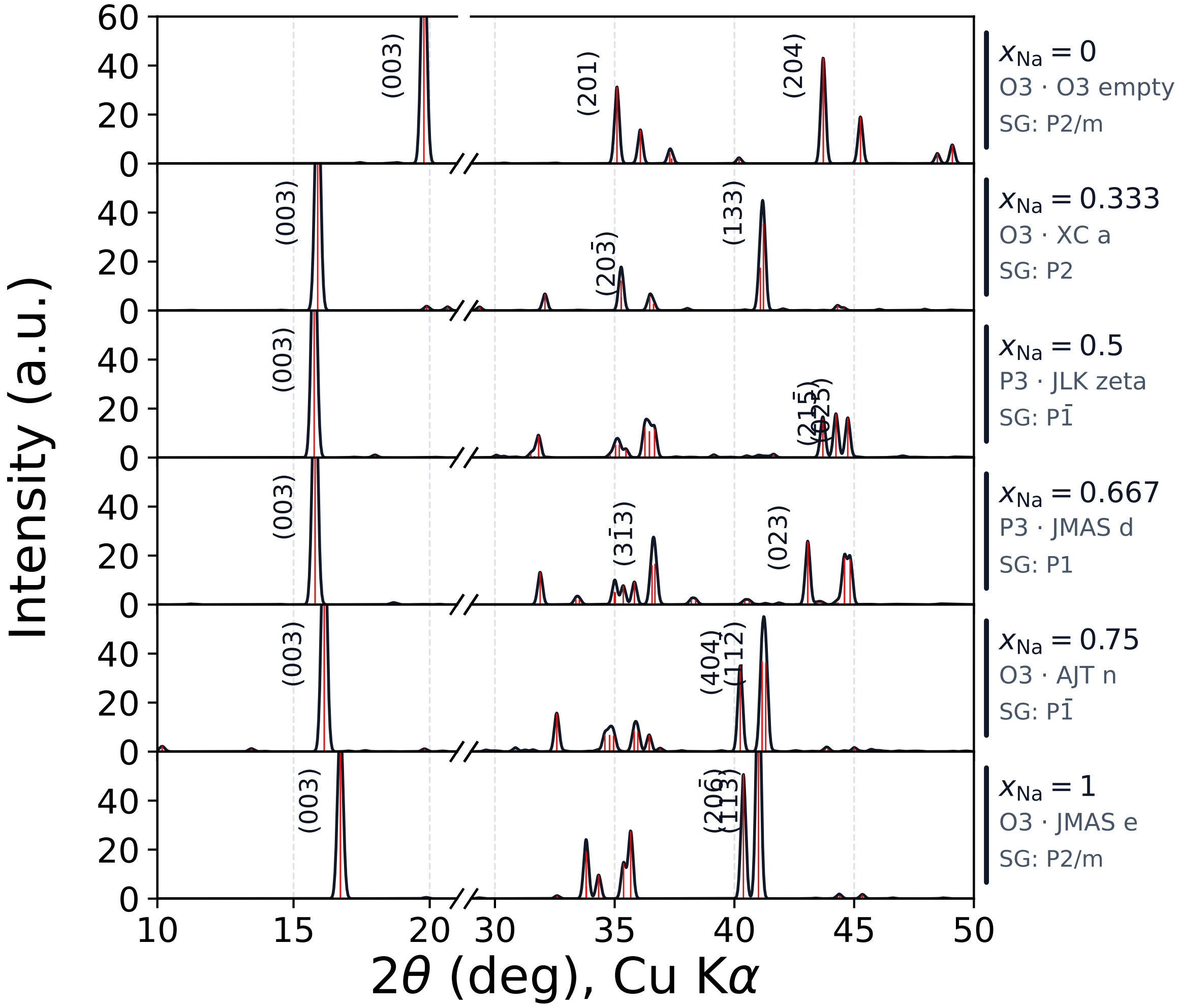

**Figure S7 (continued).** (**b**) $Na_xTi_{0.5}Fe_{0.5}O_2$.

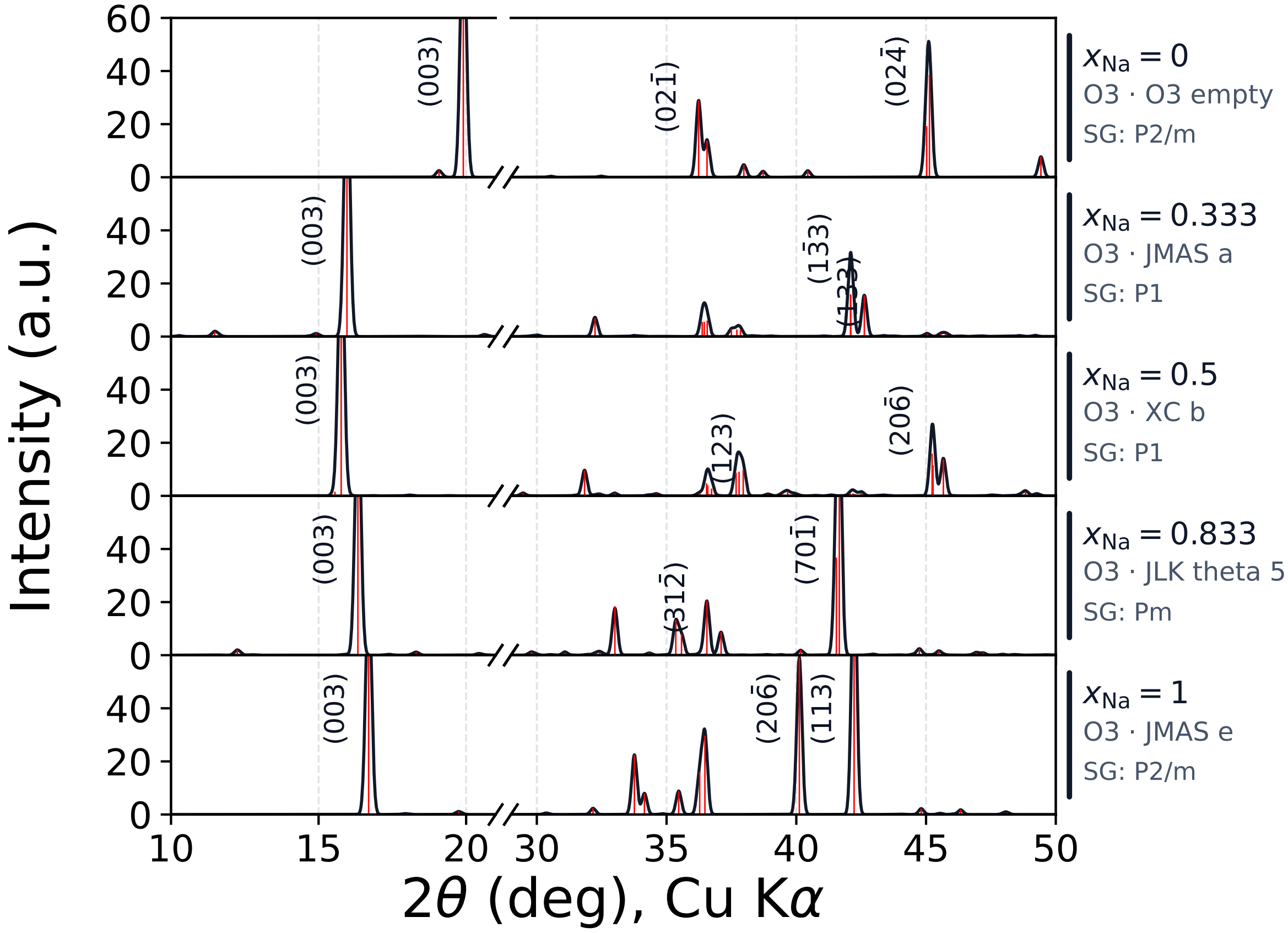

**Figure S7 (continued).** (**c**) $Na_xTi_{0.5}Co_{0.5}O_2$.

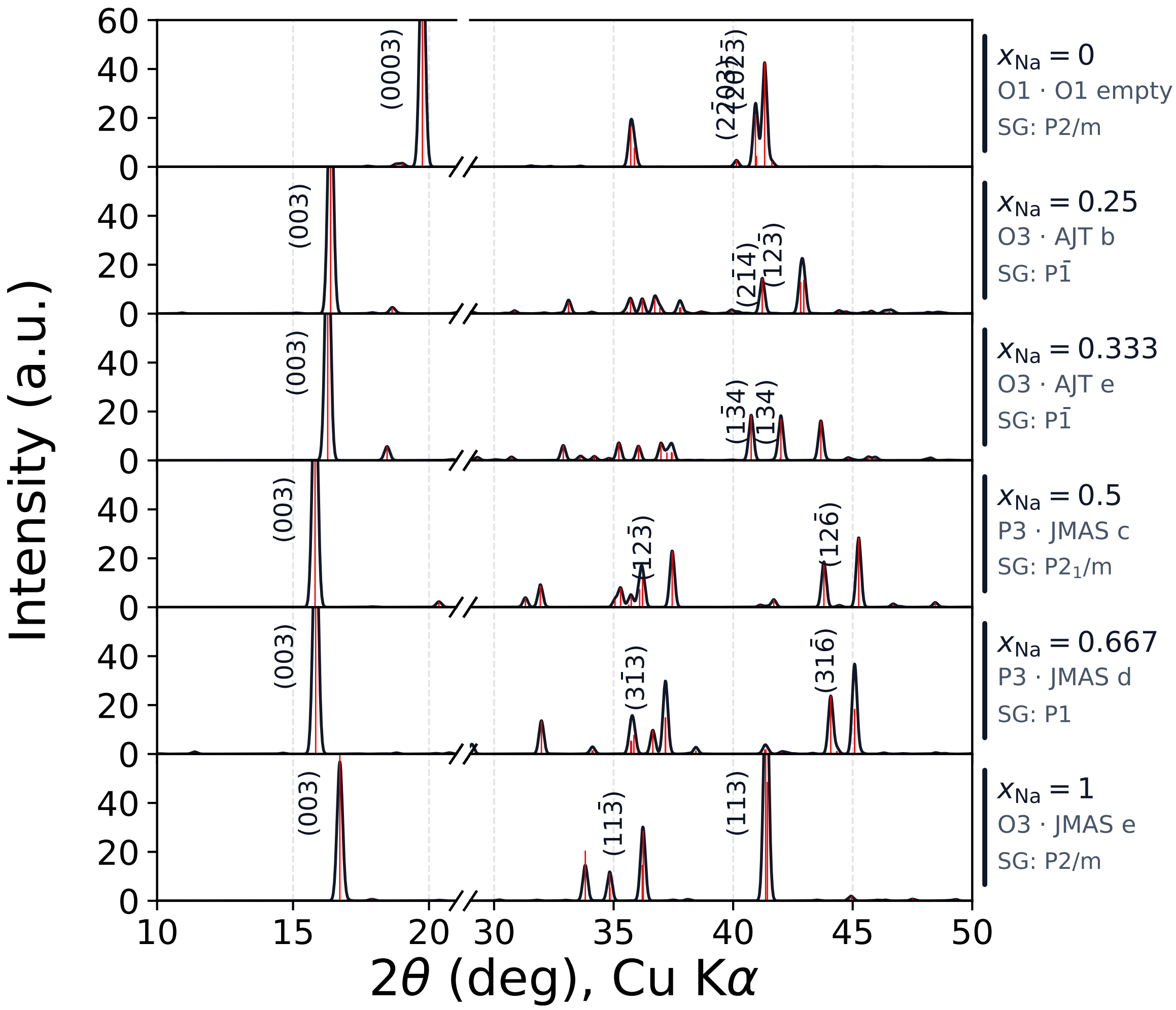

**Figure S7 (continued).** (**d**) $Na_xTi_{0.5}Ni_{0.5}O_2$.

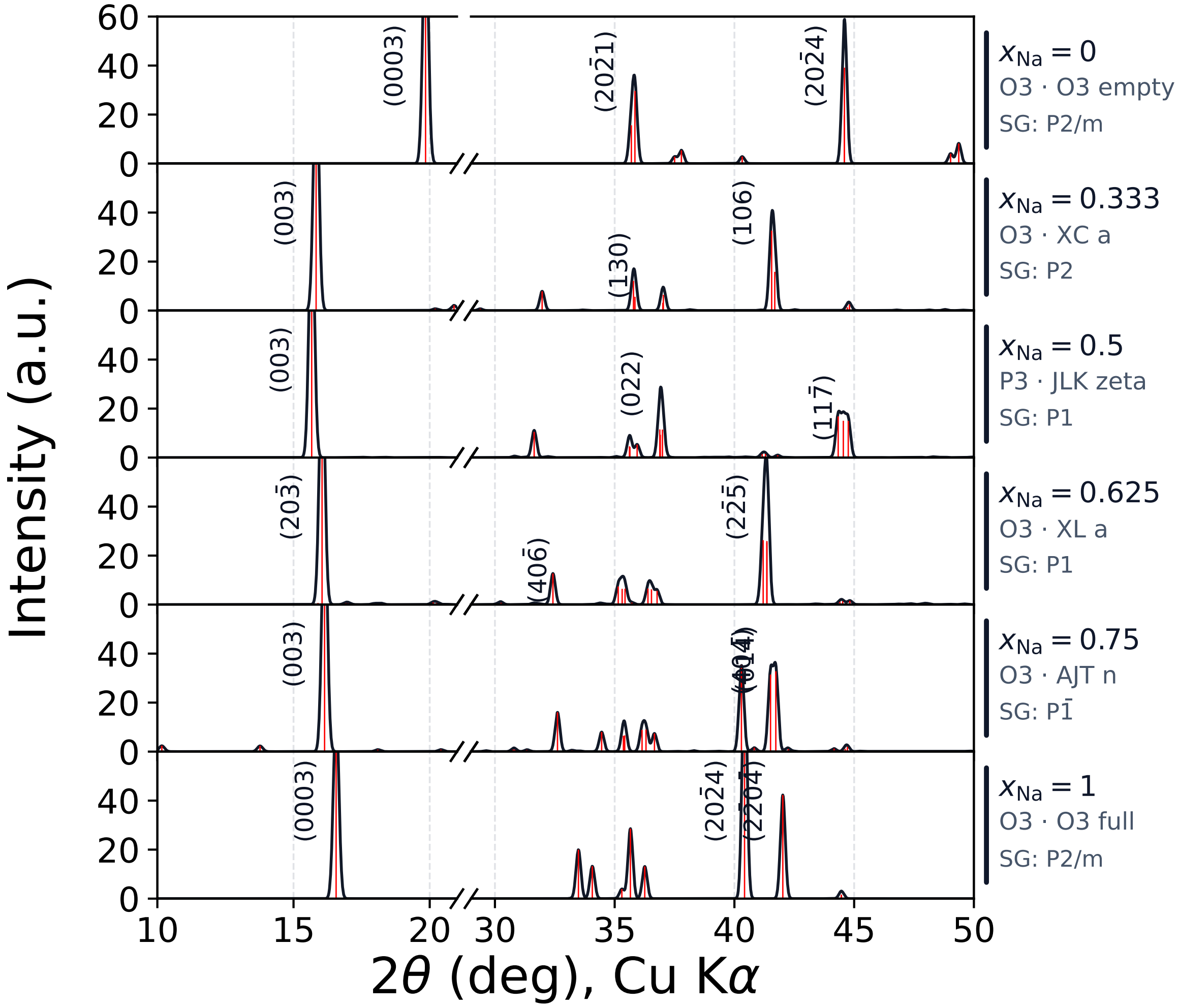

**Figure S7 (continued).** (**e**) $Na_xMn_{0.5}Fe_{0.5}O_2$.

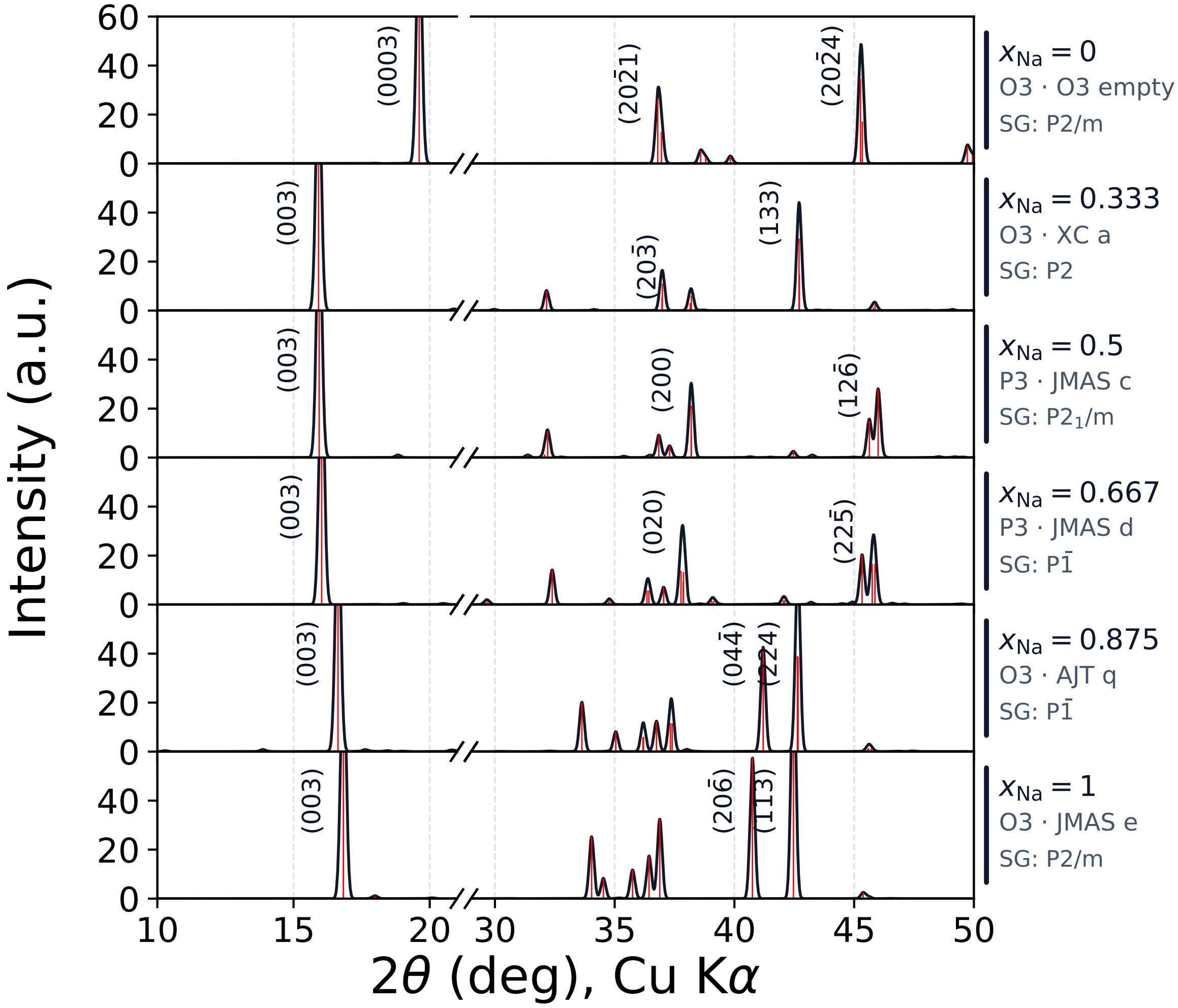

**Figure S7 (continued).** (**f**) $Na_xMn_{0.5}Co_{0.5}O_2$.

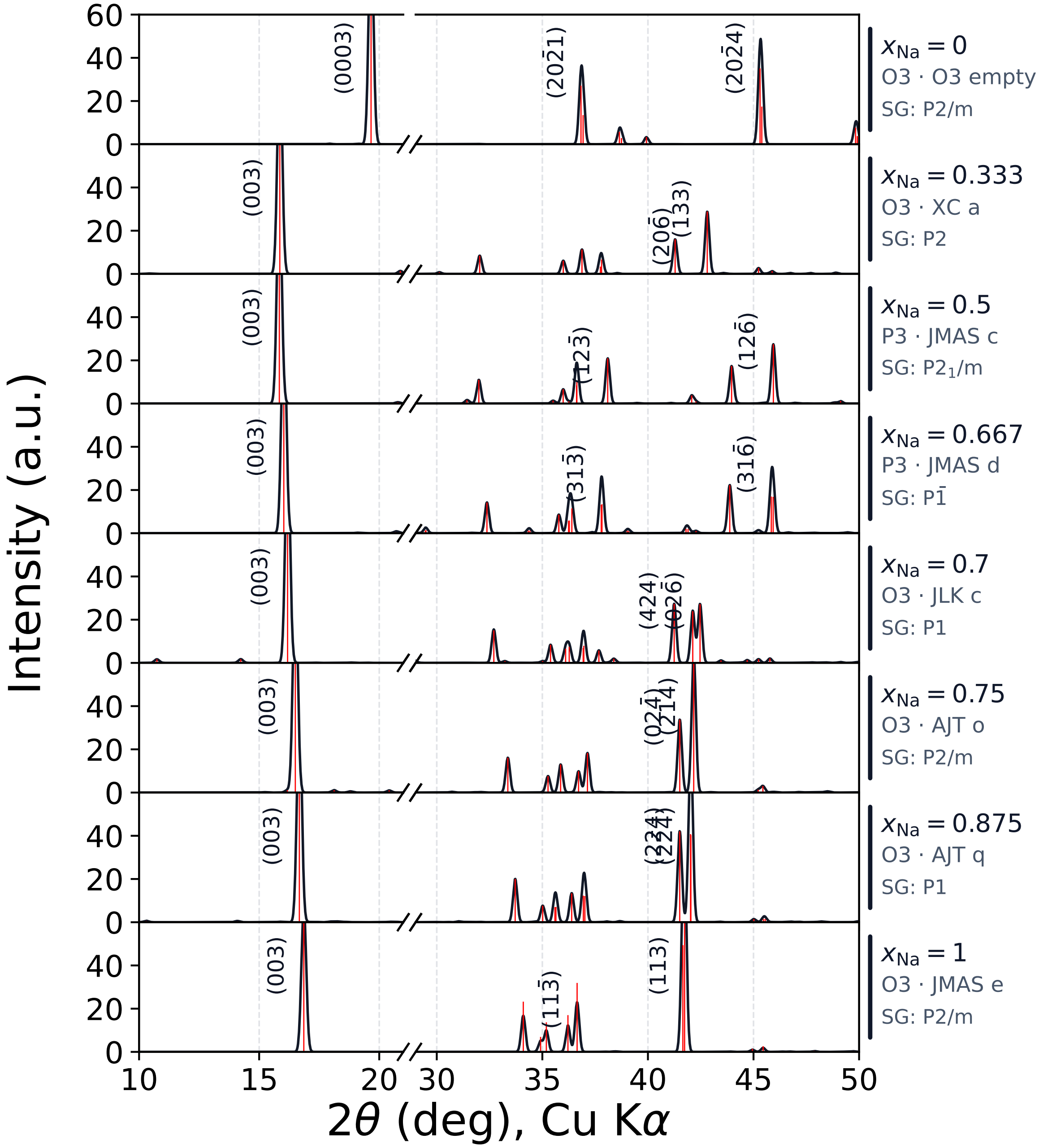

**Figure S7 (continued).** (**g**) $Na_xMn_{0.5}Ni_{0.5}O_2$.

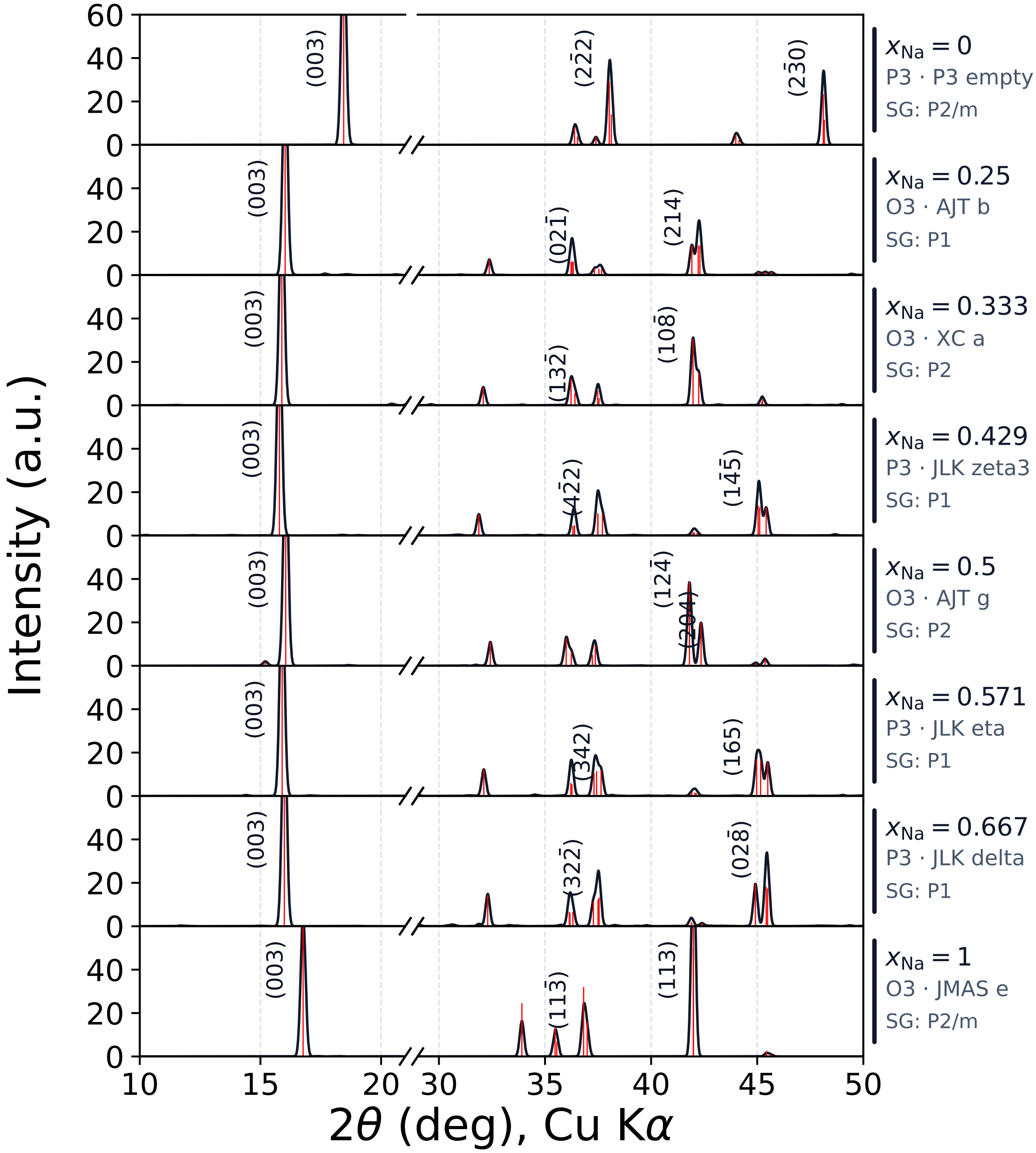

**Figure S7 (continued).** (**h**) $Na_xFe_{0.5}Co_{0.5}O_2$.

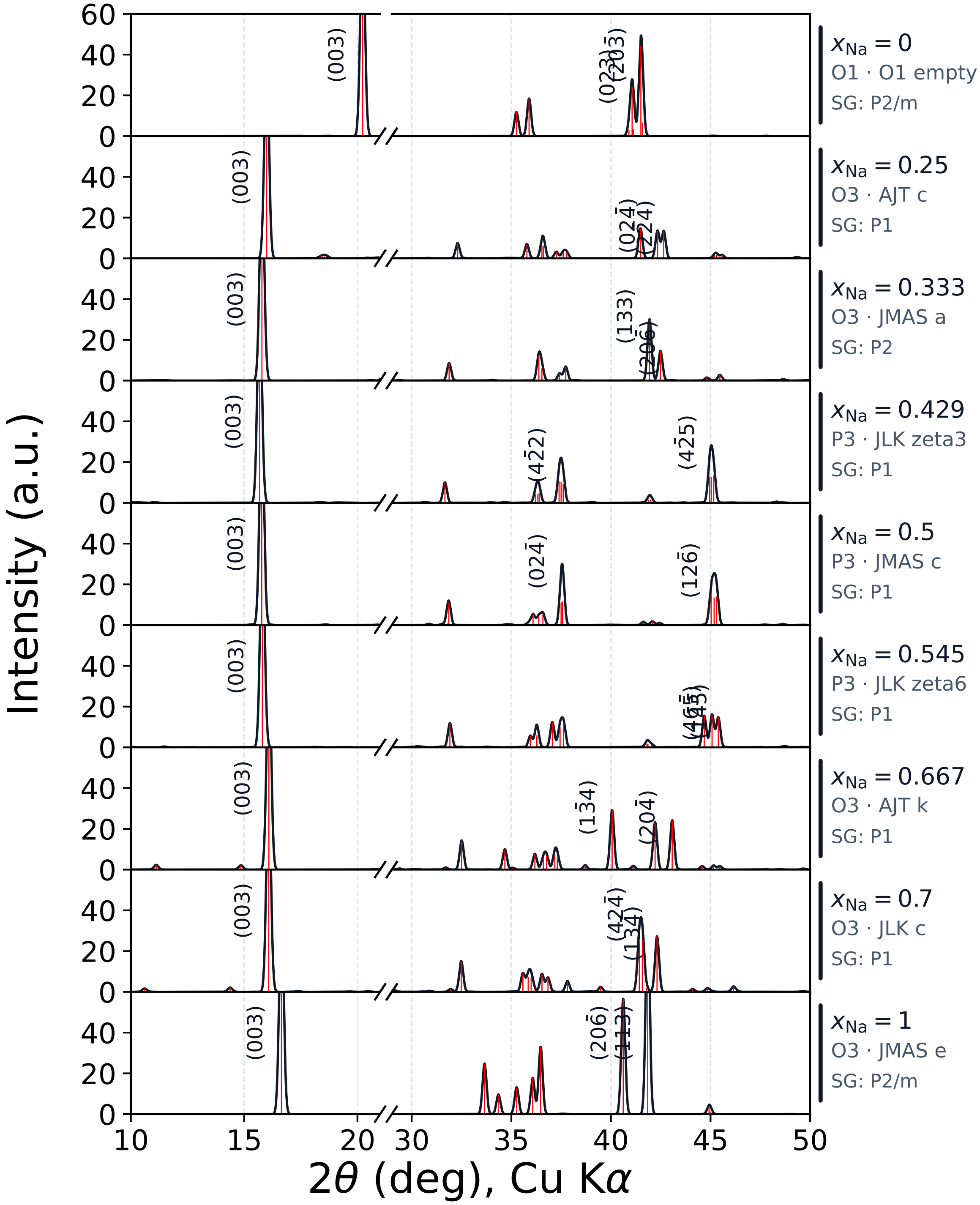

**Figure S7 (continued).** (**i**) $Na_xFe_{0.5}Ni_{0.5}O_2$.

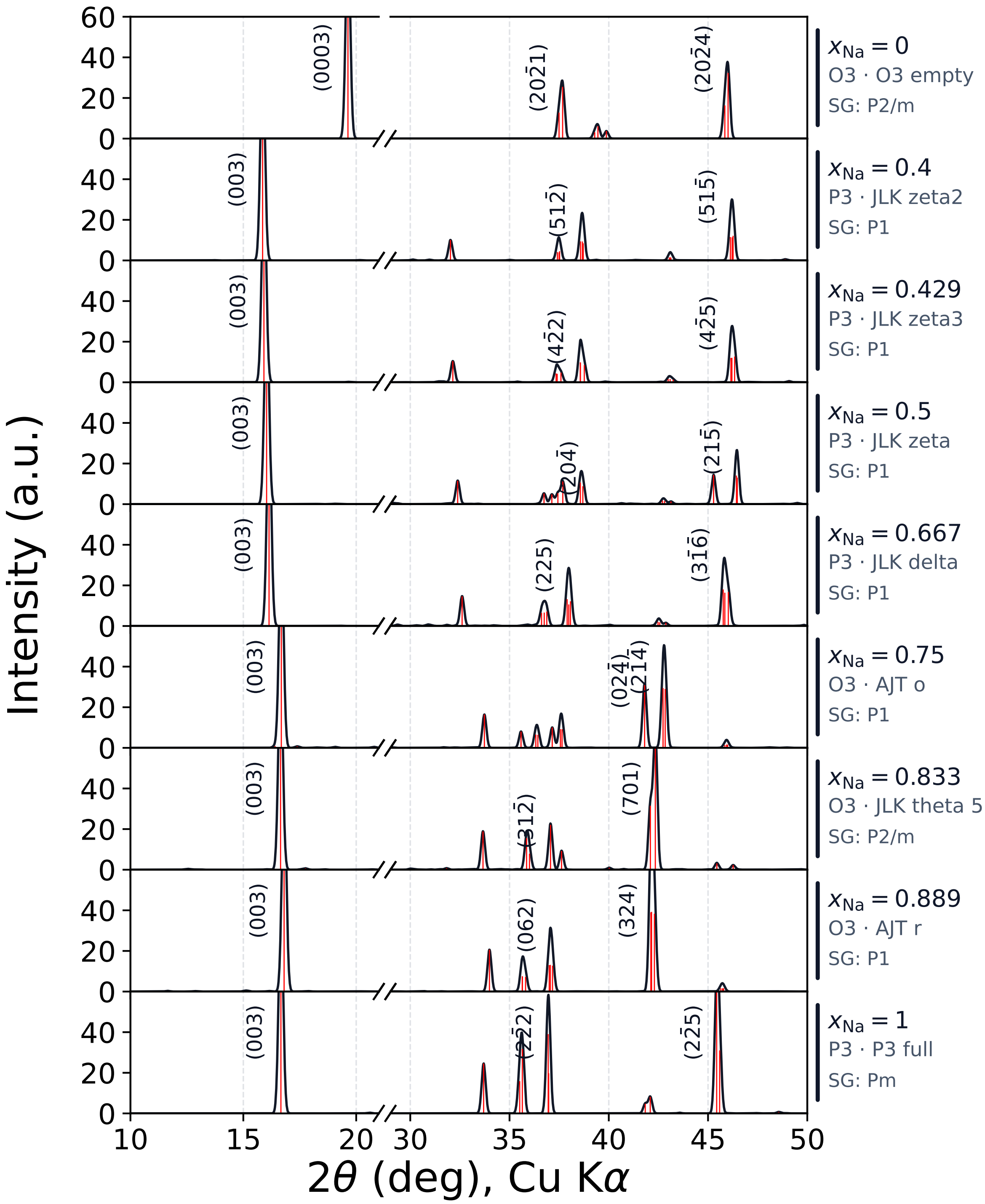

**Figure S7 (continued).** (**j**) $Na_xCo_{0.5}Ni_{0.5}O_2$.